\documentclass[iop]{emulateapj}
\usepackage{natbib}
\usepackage{txfonts}
\usepackage{graphicx}
\usepackage{wasysym}

\newcommand{\ha}{H$\alpha$}
\newcommand{\bmv}{B$-$V}

\newcommand{\hafuv}{F$_{\rm H\alpha}/$F$_{\rm FUV}$}
\newcommand{\lgflux}{$\log(\rm F_{\rm H\alpha})$}

\newcommand{\sbunit}{ergs s$^{-1}$ cm$^{-2}$ arcsec$^{-2}$}
\newcommand{\fluxunit}{ergs s$^{-1}$ cm$^{-2}$}

\shortauthors{Watkins et al.}

\begin{document}

\title{Star-Forming Environments Throughout the M101 Group}
  \shorttitle{M101 \ion{H}{2} Regions}

\author{Aaron E. Watkins\altaffilmark{1,2}, 
  J. Christopher Mihos\altaffilmark{1},
  Paul Harding\altaffilmark{1}
  }

\altaffiltext{1}{Department of Astronomy, Case Western Reserve
  University, Cleveland, OH 44106, USA}
\altaffiltext{2}{Astronomy Research Unit, University of Oulu,
  FIN-90014, Finland}

\begin{abstract}

We present a multiwavelength study of star formation within the nearby
M101 Group, including new deep \ha \ imaging of M101 and its two
companions.  We perform a statistical analysis of the \ha \ to FUV
flux ratios in \ion{H}{2} regions located in three different
environments: M101's inner disk, M101's outer disk, and M101's lower
mass companion galaxy NGC~5474.  We find that, once bulk radial trends
in extinction are taken into account, both the median and scatter in
\hafuv \ in \ion{H}{2} regions are invariant across all of these
environments.  Also, using Starburst99 models, we are able to
qualitatively reproduce the distributions of \hafuv \ throughout these
different environments using a standard Kroupa IMF, hence we find no
need to invoke truncations in the upper mass end of the IMF to explain
the young star-forming regions in the M101 Group even at extremely low
surface density.  This implies that star formation in low density
environments differs from star formation in high density environments
only by intensity and not by cloud-to-cloud physics.

\end{abstract}

\keywords{galaxies:individual(M101), galaxies:individual(NGC5474),
  galaxies:evolution, galaxies:spiral, galaxies:star formation}

\newpage

\section{Introduction}
The extended, low surface brightness (LSB) outer disks of galaxies are
a poor fit to idealized models of galaxy formation theory.  Absent
extenuating circumstances, $\Lambda$CDM predicts that galaxies form
``inside-out'', hence are youngest at their largest radii.  Yet real
galaxies' often smooth, red outer isophotes imply the opposite
\citep[e.g.][]{bakos08, zheng15, laine16}.  In fact, old red giant
branch (RGB) stars typically have longer scale lengths than main
sequence stars \citep[e.g.][]{davidge03, vlajic09, vlajic11}, and any
young stars present in outer disks tend to be sparsely distributed
\citep[e.g.][]{barker07, davidge10}.  Outer disks are not simply an
LSB continuation of inner disks.

Star formation is also inefficient in outer disks, with gas
consumption timescales exceeding a Hubble time \citep{thilker07,
  bigiel10}.  This is similar to LSB galaxies
\citep[e.g.][]{mcgaugh94a, burkholder01, boissier08}, suggesting that
star formation physics changes in low density environments.  Jeans
stability criteria suggest that low gas column density results in
depressed or truncated star formation \citep[with an apparent
  threshold below around $\Sigma_{HI} \sim 10^{20}$--$10^{21}$
  cm$^{-2}$, e.g.][]{hunter86, skillman87, vanderhulst87}, but star
formation may also be suppressed on large scales via dynamically
induced stability \citep[e.g.][]{zasov88, kennicutt89}. The latter
suggests disks should have a star formation truncation radius
\citep{martin01}, with star formation taking place beyond this only in
local high density pockets \citep[e.g.][]{courtes61, ferguson98a,
  gildepaz05, thilker05}.

Despite its scarcity and inefficiency, this in situ outer disk star
formation could fully account for all of the outer disk stellar mass
in some galaxies \citep[depending on the star formation history,
  SFH;][]{zaritsky07}.  However, outer disk star formation is present
in only $\sim$4\%--14\% of star-forming galaxies out to $z=0.05$
\citep{lemonias11}, hence it may not be sufficient to explain outer
disk formation in general.  It also may not be necessary: many authors
have proposed that much outer disk stellar mass can be accounted for
through radial migration \citep{sellwood02, debattista06}, which can
migrate early generations of inner disk stars outward via resonances
with transient spiral arms, bars, or couplings thereof
\citep[e.g.][]{roskar08, sanchez09, schonrich09, minchev11,
  roskar12}.

Because our empirical star formation laws \citep[e.g. the conversion
  of \ha \ flux to star formation rate, SFR;][]{kennicutt94} were
derived in high density environments, accounting for the fraction of
stellar mass that formed in-situ in outer disks assumes that these
laws remain unaltered in low density environments.  If this is not
true, conclusions drawn from typical star formation indicators about
gas consumption timescales, star formation efficiency, and so on will
be erroneous in outer disks and other similar environments.  Consider,
for example, two star-forming regions of equal mass and age, and so
equal in predicted SFR.  H$\alpha$ emission is sensitive to the initial mass
function \citep[IMF, e.g.][]{sullivan04}; hence, if one region lacks
massive O stars, it will emit fewer ionizing photons, resulting in
lower H$\alpha$ flux.  Measuring its SFR using a standard
H$\alpha$--SFR conversion factor will thus underestimate its true SFR.

It remains an open question if star formation physics changes in low
density environments.  Whether or not such a change occurs depends on
whether or not changes in the underlying structure of the
disk---surface mass density, gas velocity dispersion, gas phase,
turbulence, etc.---affects the formation and subsequent evolution of
molecular clouds and star clusters.  For example, \citep{meurer09}
argue that the formation of dense bound clusters is inhibited in
regions of low mass surface density because the midplane pressure in
the disk influences internal cloud pressures \citep[see
  e.g.][]{dopita03}.  If massive stars form via competitive accretion
\citep{larson73}, in which interactions between protostars drive mass
segregation and subsequent gas accretion in high density cluster
cores, protostars in low-density clusters would suffer fewer
interactions and accrete less mass, inhibiting the growth of high-mass
stars \citep[e.g.][]{bonnell04}.  Seeking out changes to the IMF in
populations of young clusters could thus help determine how sensitive
star formation within dense cores and molecular clouds is to the
surrounding environment.

Some evidence does indicate that the cloud-to-cloud physics of star
formation may be influenced by the local surface density of the disk.
In inner disks, star formation follows a power law of the form
$\Sigma_{\rm SFR} \propto \Sigma_{\rm gas}^{\alpha}$ with the measured
value of $\alpha$ ranging between $\sim$1 and 1.5 \citep[as originally
  proposed by Schmidt 1959, and subsequently confirmed
  observationally, e.g.][]{kennicutt89, kennicutt98, kennicutt07,
  bigiel08}.  Such studies have been much rarer in outer disks and
other low density environments, partly because of lack of CO emission
\citep[likely due to low metallicity or changes in ISM
  pressure;][]{elmegreen15}.  However, those that have broached this
regime find a significantly steeper value of $\alpha$
\citep[$\sim$2--3;][]{bigiel08, bigiel10, bolatto11, schruba11},
implying a significantly less efficienty coupling between star
formation and gas density than found in the inner disk.

Clues to this difference may come from dwarf irregular (dIrr) or LSB
galaxies, which, like outer disks, are often gas-dominated and low in
mass surface density \citep{mcgaugh97, vanzee97, hunter11}.  Stellar
and gaseous disks in dIrr galaxies are also thicker than normal
spirals \citep{elmegreen15}, which can help stabilize them
\citep{vandervoort70}; outer disks may again be similar, as they are
frequently warped \citep{sancisi76, vanderkruit87, bottema87,
  garcia02, vaneymeren11}.  In a case study of the dIrr Sextans A,
\citet{hunter96} found that stars still form at a slow rate in the
peaks of the gas distribution even though dynamical arguments suggest
this should not be the case \citep[e.g.][]{toomre64, kennicutt89}.
\citet[][]{vanzee97} found similar results for six additional LSB
dwarf galaxies.  These galaxies lack interaction signatures, hence
\citet{vanzee97} proposed that star formation therein is likely
regulated by feedback, such as stellar winds or supernovae, locally
compressing gas.  Such a mechanism may be necessary to sustain star
formation in environments that lack the periodic forcing provided by
spiral arms or bars, which may also be absent in outer disks
\citep{watkins16}.

One might thus consider whether these differing mechanisms yield
observationally distinct populations of young clusters and \ion{H}{2}
regions.  This is currently a topic of considerable discussion, and
some previous studies have uncovered hints to this effect.
\citet{hoversten08}, for example, found that integrated colors of
dwarf and LSB galaxies suggest a deficiency in high-mass stars; this
may be related to their low integrated SFRs \citep{gunawardhana11}.  A
lack of high-mass stars may also account for the lack of
high-luminosity \ion{H}{2} regions in dwarfs and LSB galaxies
\citep{helmboldt05, helmboldt09}.  Yet \citet{schombert13} found that
when all 54 LSB galaxies in their sample were taken as a whole, the
\ion{H}{2} region luminosity function (LF) was the same as that found
in normal spirals, hence the lack of bright \ion{H}{2} regions in LSB
galaxies could be merely a sampling effect given the intrinsic
rareness of high-luminosity \ion{H}{2} regions in general.

One means of informing this debate is to compare and contrast
different star formation tracers.  SFR conversion factors assume the
following: that stars are sampled from a universal IMF, that the SFH
is constant over Gyr timescales, and that there is no attenuation by
dust \citep{kennicutt83, donas87}.  Under those assumptions, different
SF indicators should yield identical SFRs.  Conversely, if
different SF indicators yield different SFRs, one or more of those
assumptions must be invalid.  For example, when properly accounting
for dust, \ha \ emission traces mainly O stars with masses $M_{*}
\gtrsim 10M_{\odot}$, while far ultraviolet (FUV) emission traces O
and B stars down to $M_{*} \sim 3M_{\odot}$ \citep{kennicutt12};
hence, variation in the \ha \ to FUV flux ratio (hereafter \hafuv) can
be used to study the behavior of the high mass end of the IMF in young
clusters \citep[e.g.][]{lee09}.

This ratio also shows trends that may hint at environmentally
dependent star formation physics: globally averaged \hafuv
\ correlates with galaxy stellar mass \citep{boselli09, lee09}, with
\emph{R} band surface brightness \citep[but see Weisz et
  al. 2012]{meurer09}, and with radius in some galaxies
\citep{thilker05, goddard10, hunter10}.  Unfortunately, \hafuv \ is
sensitive to a large number of variables, which makes interpretation
of these trends difficult.  In addition to dust extinction \citep[in
  fact, \hafuv \ correlates extremely well with extinction, to the
  point that it can itself be used as an extinction
  estimator;][]{cortese06, koyama15}, \hafuv \ decreases rapidly with
age \citep[e.g.][]{leroy12} as the high-mass stars traced by \ha
\ emission die off.  IMF sampling effects play a similar role, and
introduce stochasticity in \ha \ emission at low \ion{H}{2} region
mass, where a given \ion{H}{2} \ region may be powered by a single O
or B star \citep{lee09, lee11}.  These degeneracies have led to much
discussion regarding the true origin of the observed \hafuv \ trends,
with explanations ranging from a changing IMF at low density
  \citep{pflamm08, meurer09, pflamm09}, to age effects
\citep{alberts11}, to stochastic sampling \citep{goddard10,
  hermanowicz13} or non-uniform SFHs \citep{weisz12}.

The nearby face-on spiral M101 (NGC~5457) provides a unique target for
investigating the connection between star formation and
local environment.  Broadband imaging by \citet{mihos13} found
extremely blue (\bmv$\sim$0.2--0.4) colors in the extended LSB outer
disk of the galaxy, implying a significant population of young stars
at large radius.  This is also apparent from deep \emph{GALEX} FUV and
near ultraviolet (NUV) imaging, which show that the galaxy has an XUV
disk \citep{thilker07}.  Given its disturbed morphology, this
extended star formation likely resulted from an interaction with one
or both of its companions, NGC~5477 and NGC~5474 \citep{mihos13}.
Both companions are star-forming themselves and nearby on the sky.
The M101 galaxy group thus provides examples of three different kinds
of star-forming environments in close proximity; a high-mass
star-forming disk, an LSB star-forming outer disk, and two
star-forming companion galaxies with lower mass.

As such, we targeted the M101 Group for deep narrow-band \ha \ imaging
with the Burrell Schmidt Telescope at Kitt Peak National Observatory
(KPNO).  The Burrell Schmidt's wide field of view allows for a direct
comparison of all three galaxies in the M101 Group in a single mosaic
image.  We use our \ha \ narrow-band imaging data in conjunction with
the deepest available \emph{GALEX} FUV and NUV images of M101 and its
companions in order to investigate the statistical properties of the
\hafuv \ ratio in both the \ion{H}{2} regions as a function of these
three environments.  In Section 2, we give a brief overview of our
observation and data reduction procedures.  In Section 3, we describe
our methodology for analysing the \ion{H}{2} regions, including
extinction correction, \ion{H}{2} region selection, and photometry.
We present the results of these analyses in Section 4.  In Section 5,
we discuss the implications of our results in the context of previous
analyses of the \hafuv \ ratios of galaxies, as well as the broader
applicability of our results.  We conclude with a summary in Section
6.

\section{Observations and Data Reduction}

Here we present a discussion of our observing strategy and data
reduction techniques.  We briefly review these here; for an exhaustive
description, we refer the reader to our previous work \citep[and
  references therein]{watkins14, mihos17}.  However, this previous
work used broadband filters, hence we focus in this section on
adjustments to these procedures necessary in shifting to narrow-band
imaging data.

\begin{figure*}
  \centering
  \includegraphics[scale=0.32]{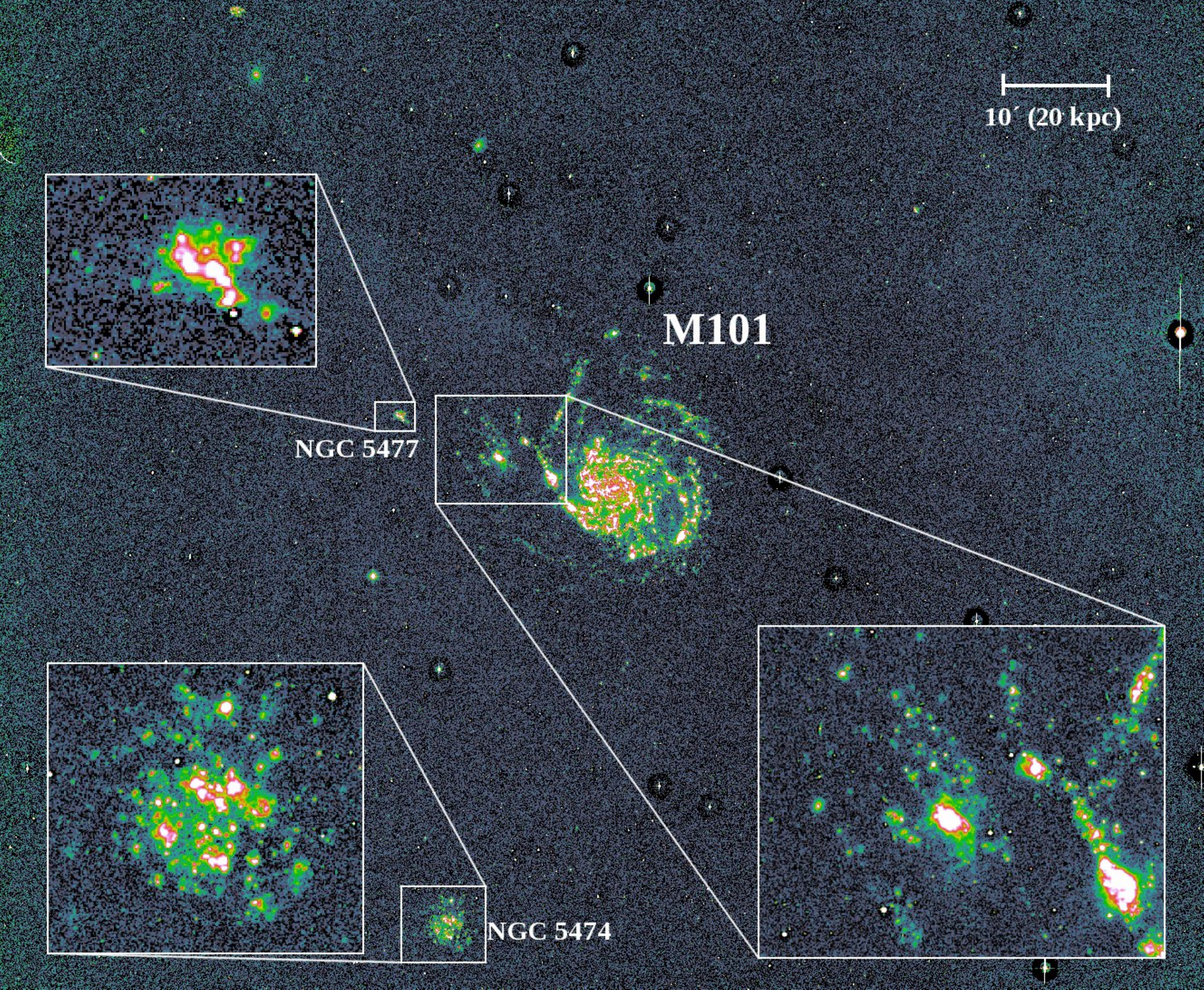}
  \caption[\ha \ difference image mosaic of the M101 Group]{A view of our
    difference image mosaic, showing \ha \ emission in M101 and its
    companions.  Insets are shown of NGC~5477, NGC~5474, and the
    eastern side of M101 containing the giant \ion{H}{2} region
    complexes NGC~5471 (center frame) and NGC~5462 (at the lower
    right), to showcase the wealth of low surface brightness structure
    we detect.  Pixels saturate (white) in this image at
    $\sim$2.85$\times 10^{-16}$ \sbunit. North is up and east is to
    the left.
    \label{fig:diffim}}
\end{figure*}

\subsection{Observations}

We observed M101 with the Burrell Schmidt telescope at KPNO in spring
of 2014, using two custom narrow-band interference filters. The two
filters have central wavelengths at 6589~\AA \ and 6726~\AA
\ (hereafter the on-band and off-band filters, respectively), with
$\sim$100~\AA \ widths, necessitated by the Schmidt's fast f$/3.5$
beam.  The on-band covers \ha \ at M101's velocity \citep[$\sim$240 km
  s$^{-1}$;][]{devau91}, while the off-band filter covers the adjacent
stellar continuum; given M101's low inclination, all \ha \ emission
from the galaxy lies within a region of the on-band filter with
$\sim$96\% transmission.  The on-band filter bandpass is wide enough
to include Milky Way emission; however, M101 is located at a high
Galactic latitude in a field relatively free of Galactic cirrus
\citep{schlegel98, schlafly11}, limiting contamination.  We observed
only on moonless, photometric nights, using exposure times of 1200 s
for both filters, with dithers of $\sim$0\fdg5 between exposures
to remove large-scale artifacts such as flat-fielding errors and
scattered light.  This resulted in sky levels of 200--300 ADU in the
on-band filter, and 150--250 ADU in the off-band.  In total, we
observed M101 in each filter for $71 \times 1200$ s (nearly 24 hours
per filter).

Due to low sky counts in the narrow-band filters, we could not
construct flats from night-sky frames alone.  To construct the flats,
we started with twilight exposures; however, given our large field of
view, these twilight flats contained noticeable gradients induced by
the setting sun.  We therefore also produced flats without gradients
using offset night-sky frames with exposure times equal to our object
frames (1200 s for both filters), as we did in constructing flat
fields for our broadband imaging \citep[see][]{watkins14, mihos17}.
The final twilight flats consisted of $\sim$110 individual exposures
per filter, averaging $\sim$20000 ADU px$^{-1}$, while final night sky
flats totaled $82 \times 1200$ s exposures in the on-band, and $74
\times 1200$ s exposures in the off-band.  We defer a discussion of
how we used both of these flats for the final reduction to the next
section.

Finally, we observed spectrophotometric standard stars from
\citet{massey88} for photometric calibration, along with several 1200
s exposures of Arcturus in order to model internal reflections and the
extended wings of the Schmidt point-spread function \citep[PSF;
  see][]{slater09}.

\subsection{Data Reduction}

We began data reduction by applying a standard overscan and bias
subtraction, correcting for nonlinear chip response, and applying a
WCS to each frame.

Flat-fielding took place in stages.  We first constructed master
twilight flats by median-combining all $\sim$110 twilight exposures
per filter.  To remove gradients in the twilight flats, we then
constructed night-sky flats as described in previous works
\citep{watkins14, mihos17}. In short, for each frame, we created an
initial mask using the IRAF\footnote{IRAF is distributed by the
  National Optical Astronomy Observatory, which is operated by the
  Association of Universities for Research in Astronomy (AURA), Inc.,
  under cooperative agreement with the National Science Foundation.}
task \emph{objmask}, hand-masked any remaining artifacts (typically
light scattered by stars just off-frame), and combined the resulting
masked frames into a preliminary flat.  We then flattened and
sky-subtracted all night-sky frames using this preliminary flat,
combined the flattened and sky-subtracted images into a new flat, and
repeated for 5 iterations, until the flat field converged.

We isolated the twilight flat gradients through division by the
gradientless night-sky flats.  We then modeled and divided the planes
out of the twilight flats, resulting in final generation flat
fields.  This is mathematically equivalent to using the night-sky
flats (modulo uncertainty in the gradient fits), but with the improved
Poisson statistics of the twilight flats on small scales.

Mild fringing is visible in all of our on-band images at an amplitude
of $\sim$0.1\%, but absent in the off-band images.  As M101 is far
from the ecliptic plane (hence from zodial light contributions), the
main contributor of this fringing is telluric emission lines
\citep[OH;][]{massey00}, which are not present in the off-band filter.
We thus measure and correct for fringing in on-band frames only.
Because scattered sunlight dominates the telluric emission in the
twilight frames, the twilight flats lack the fringe pattern.  Hence,
to isolate the pattern, we divide the night-sky flat (which does
contain the pattern) by the twilight.  We then scale a normalized
version of this fringe map to the sky level of each on-band frame
(corrected for large-scale gradients) and subtract it from each frame.
Because this fringing is present on all on-band night-sky frames, we
reconstruct the on-band night-sky flat after fringe removal and
rederive the on-band twilight flat gradient before flat-fielding the
on-band object frames.

For our final flux calibration, we observed spectrophotometric
standard stars from the \citet{massey88} catalog.  We derived
photometric zero points by convolving our filter transmission curves
over the spectra of these stars to derive filter magnitudes (defined
as $-2.5 \log(F_{\rm filt})$ for simplicity, where $F_{\rm filt}$ is
the total flux in ADU of the star through the filter), which we
compared with instrumental magnitudes derived through photometry of
each exposure of each star.  In each observing run, we observed 12
unique standard stars, several of which we observed multiple times to
improve the final zero points.  Due to uncooperative weather, we did
not achieve adequate airmass coverage from these standard star
observations; instead, we derived airmass terms for each filter using
photometry of SDSS DR8 \citep{aihara11} stars found in the individual
exposures of M101 (this is described in more detail below).  The
photometric zero points are thus simply:

\begin{equation}
  ZP = -2.5\log(F_{\rm filt}) - (m_{\rm inst} - \kappa \sec z)
\end{equation}

where $m_{\rm inst}$ is the instrumental magnitude and $\kappa$ is the
airmass term.

For each filter, we take as the zero point the median value of the
zero points derived from each star.  The standard error on the median
is $1.253\sigma/\sqrt{N}$, hence errors on the two filter zero points
are $\sigma_{\rm on} = 0.006$ mag and $\sigma_{\rm off} = 0.003$ mag; this
translates to an error of $\sim$2\% on M101's total flux.  In our
final mosaic of M101, 1 ADU per pixel per 1200 s is equal to an \ha
\ surface brightness of $\Sigma_{H\alpha} = 3.557 \times 10^{-18}$
ergs s$^{-1}$ cm$^{-2}$ arcsec$^{-2}$, or an emission measure of
EM$\sim$1.78 cm$^{-6}$ pc.  Using this flux calibration, we find good
agreement (to within $\sim$3\%) with the value of M101's total
flux published by \citet{kennicutt08}, measured within their value of
R$_{25}$.

To reduce scattered light artifacts, we also remove reflections and
diffuse halos around bright stars in all frames in the manner
described by \citet{slater09}.  Briefly, we use deep (1200 s) exposures
of Arcturus at different positions on the chip to measure and model
these reflections and halos, then scale and subtract them from all
stars brighter than \emph{V}$=10.5$ found in each frame.  We do this
scaling via a rough photometric calibration using SDSS stars found in
each field, assuming our on-band filter is equivalent to SDSS \emph{r}
with no color term.  This produces fairly robust scalings for the
reflection- and halo-subtraction process; only for the brightest stars
(\emph{V} $>8$) did we need to tweak the derived magnitudes by hand in
order to produce an acceptable subtraction.  Given this stability, the
large number of SDSS stars in each frame, and the improved airmass
coverage, we choose to use the airmass terms derived in this way over
those derived from the standard star exposures for our flux
calibration.  This choice has little effect on the calibration, as the
airmass terms are quite small ($\lesssim 0.1$) for both filters.

Finally, we sky-subtract each frame by masking all bright stars and
galaxies, fitting sky planes to each masked image, and subtracting
these planes from the frames.  To preserve precise flux scaling, we
then scale these images to zero airmass and median-combine them into
two final mosaics (an on-band and off-band) using the IRAF tasks
\emph{wregister} and \emph{imcombine}.  Because these two mosaics
combine many exposures taken under variable observing conditions, a
direct subtraction of the two does not produce a clean difference
image, making it difficult to identify LSB regions.  Hence we create a
third mosaic using individual pairs of images taken back-to-back.  We
align both images to within 0.1 pixels, photometrically scale and subtract
the off-band images from the on-band, and combine the individual
difference images into one mosaic, as before.  While we use this
difference mosaic to display our data, all H$\alpha$ fluxes quoted
henceforth are measured from the on-band and off-band mosaics, which
preserve the flux calibration most accurately.

\begin{figure*}
  \centering
  \includegraphics[scale=0.68]{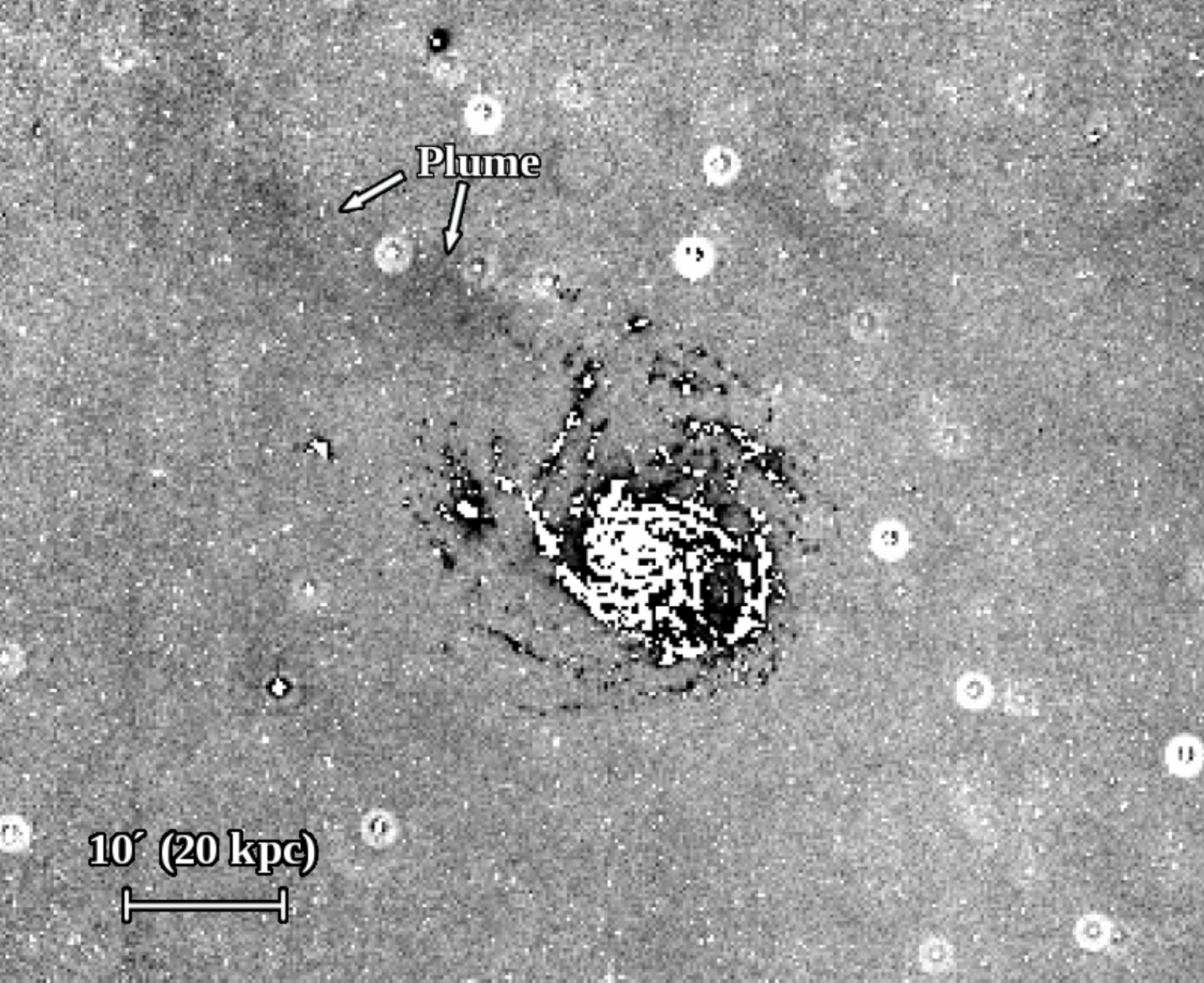}
  \caption[Tentative diffuse plume northeast of M101]{A masked, 9$\times$9
    pixel median-binned image of our difference mosaic, showcasing a
    plume of extremely diffuse H$\alpha$ emission.  North is up and
    east is to the left. \label{fig:diffuse}}
\end{figure*}

In our previous work, the background sky scatter was dominated by
unresolved sources \citep[background galaxies and foreground stars;
  see][]{rudick10}, however the grand majority of these sources have
no emission lines that fall within our two narrow-band filters and
thus cleanly subtract out.  This, combined with our large total
exposure time, results in extremely low background noise.  We
calculate the background sky uncertainty as the dispersion in the
median count levels measured in 50--100 blank apertures with radius 15
pixels (22\arcsec) chosen adjacent to the target galaxies.  Near M101,
the background scatter in the difference image is $\sigma\sim$0.15
ADU, giving a limiting depth of $\Sigma_{\rm H\alpha, \rm
  lim}\sim5.34\times 10^{-19}$ \sbunit\ (EM$\sim0.27$).  The scatter
is slightly lower near NGC~5474 ($\sigma\sim$0.13 ADU) despite it
being nearer the edge of the mosaic; this is due to the presence of
several slightly imperfectly subtracted reflections from bright stars
near M101.

Figure \ref{fig:diffim} shows a subset of our full difference mosaic,
with several areas of interest zoomed in to showcase the wealth of LSB
\ha \ emission we detect.  We also tentatively identify an extremely
extended and LSB plume of \ha-emitting gas northeast of M101.  While
barely visible in Figure \ref{fig:diffim}, we show an enhanced image
of it in Figure \ref{fig:diffuse}, which shows our difference image
masked of bright pixels (masks shown in white) and median-binned into
9$\times$9 pixel bins.

The plume spans a length of $\sim$30~kpc, and has a characteristic
surface brightness of $\Sigma_{H\alpha} = 1.4 \times 10^{-18}$
(EM$\sim$0.7), extending from the diffuse star-forming northeast plume
region discussed by \citet{mihos13}.  When compared to adjacent
background regions of similar size and shape \citep[see][]{rudick10,
  watkins14}, this surface brightness amounts to roughly a 2$\sigma$
detection.  So well removed from M101's star-forming disk, the
ionization source for this plume is unclear.  One possibility is that
it is gas ionized by the metagalactic ionizing background, however the
feature's \ha \ surface brightness is roughly an order of magnitude
higher than expected for this phenomenon \citep{vogel95}.
Additionally, we see no evidence of diffuse ionized gas \citep[DIG;
  see, e.g.,][]{reynolds90, haffner09} in the
long, low column density \ion{H}{1} feature on the opposite, southwest
side of M101 \citep{mihos12} as might be expected if the ionization
was from the metagalactic background.  A more mundane explanation
might be that the plume is diffuse \ha \ located within our own Milky
Way galaxy.  The velocity width of our filter also covers Galactic ISM
velocities, and an examination of the \ion{H}{1} data cube of
\citet{mihos12} shows copious diffuse Galactic \ion{H}{1} projected
across the M101 Group.  If this Milky Way gas is ionized, it would
show as a patchy screen of diffuse \ha \ across our image.  However,
the spatial coincidence of the \ha \ tail with the NE plume in M101's
tidally distorted outer disk, and the lack of any comparable features
elsewhere in our mosaic (which covers 2\arcdeg $\times$ 2\arcdeg),
remains intriguing.

\subsection{\emph{GALEX} data}
In order to measure the \hafuv \ ratio, as well as to correct for
extinction, we use the deepest available \emph{GALEX} FUV and NUV images of
M101 and its companion NGC~5474.  The images of M101 were taken as
part of the guest observing program in 2008 (GI3\_05), and were first
published in \citep{bigiel10}.  These images have exposure times of
$\sim$13300 s in both FUV and NUV.  The images of NGC~5474 were taken
as part of the Nearby Galaxy Survey \citep[NGS;][]{bianchi03} and have
exposure times of 1610 s in both FUV and NUV, hence are shallower than
those of M101.  We calculate all FUV and NUV fluxes directly from the
intensity maps, while we calculate photometric errors on these fluxes
as Poisson errors using the associated high resolution relative
response maps \citep[as discussed in][]{morrissey07}.  Because
  FUV fluxes are given as monochromatic fluxes, we multiply
  all FUV fluxes by the FUV filter's central wavelength in order to
  keep the ratio \hafuv \ unitless.

\subsection{Background/Foreground Contamination}
Given the width of our filters, we detect \ha \ emission from sources
at a large range of redshifts (we cover H$\alpha$-emitting sources at
10\% transmission out to $\sim$4300 km s$^{-1}$ in our on-band
filter), resulting in both background and foreground contamination.
While background spiral and elliptical galaxies are typically
resolved, hence identifiable by eye, we also find many point sources
in the difference mosaic that are not obviously associated with the
M101 Group galaxies.

We investigated the origins of these point sources using the method
described by \citet{kellar12}.  Briefly, they define a quantity
$\Delta m = m_{\rm H\alpha} - m_{R}$, where $m_{\rm H\alpha}$ is the magnitude
of a source in their filters targeting \ha \ emission and $m_{R}$ is
the magnitude of the same source in their continuum \emph{R} band
filter, scaled such that $\Delta m = 0$ for sources with no emission
present in the \ha \ filter.  They label unresolved sources with
$\Delta m < 0$ ``\ha \ dots'', which are simply point sources that are
bright in their difference images.  As we use a narrow-band continuum
filter instead of \emph{R}, in our case $\Delta m = m_{\rm on} -
m_{\rm off}$.  We utilize the same cutoff limit as \citet{kellar12}
for ``dot'' selection, that being sources with emission line
equivalent widths $\gtrsim 30$\AA.  This corresponds to $\Delta m
\lesssim -0.3$ for our filter widths of 100\AA.

While \citet{kellar12} obtained follow-up spectroscopy of the \ha
\ bright point sources in their fields, such follow-up is beyond the
scope of our project.  Hence, we investigated the \ha \ dots in our
field by cross-referencing them with SDSS and plotting their $g-r$
vs. $r-i$ colors.  We find that the majority of the \ha \ dots in our
final mosaic lie in the region of color-color space occupied by M
stars \citep[Figure 1 of][]{finlator00}, while only a select few have
colors bluer than this.  This M star contamination results from the
width and placement of our filters; typical M star spectra contain
broad TiO absorption features, and our on-band filter's central
wavelength ($\sim$6600\AA) happens to often lie on a peak in between
two such features, while our off-band filter ($\lambda_{cen}
\sim$6700\AA) lies in an adjacent trough.  This gives M stars the
appearance of an emission line source in the difference mosaic.

Thankfully, these stars are readily identifiable as being bright in
the difference mosaic but strongly lacking in FUV emission, as well as
through available SDSS photometry.  We hence reject all sources with
\hafuv $>$ -1.4 \citep[this cutoff is also justified by Starburst99
  models, which never reach \hafuv \ higher than
  this;][]{leitherer99}, $g-r > 1.2$, and $r-i > 0.8$.  The handful of
dots with bluer colors are likely unresolved background galaxies,
unresolved star-forming dwarfs near M101, or intergalactic \ion{H}{2}
regions \citep{kellar12}.  For example, SDSS spectra of two of the
sources shows that they are quasars at $z=1.34007$
($\alpha=$211\fdg8225, $\delta=$53\fdg75559) and at $z=1.34536$
($\alpha=$211\fdg13981, $\delta$=53\fdg40635); we detect redshifted Mg
emission from both of these sources.  These bluer sources are rare,
however (we find 8 across our entire field of view, for a surface
density of $\sim$2 per square degree), hence have a negligible effect
on our analyses.

\section{Methods}
We present here our analysis of \ion{H}{2} regions in the M101 Group.
We begin by discussing our extinction correction method, then we
discuss how we identify \ion{H}{2} regions against the DIG background,
and conclude with the results of this analysis.

\subsection{Extinction Correction}
Given that we focus much of this study on the ratio \hafuv, the
components of which are separated by some $\sim$5000\AA \ in
wavelength, some manner of extinction correction is called for.
Ideally, this would be done using direct tracers of nebular extinction
such as the Balmer decrement (the H$\alpha$/H$\beta$ flux ratio).
While Balmer decrements have been published for $\sim$200 of the
brighter \ion{H}{2} regions in M101 \citep{scowen92}, we need an extinction
correction we can apply across the entire dataset, and so we choose to
employ the \emph{GALEX}-calibrated radial IRX-$\beta$ extinction correction
method described by \citet[in their Section 3.6]{goddard10}.  We recap
this method briefly here.

IRX-$\beta$ is an empirical relationship between the ratio of the
infrared and UV luminosities (the infrared excess, IRX) and the slope
of the UV continuum ($\beta$).  It works under the assumption that all
of the non-ionizing UV radiation that is absorbed by intervening dust
is reprocessed into the IR \citep{heckman95, meurer95, meurer99}.
IRX-$\beta$ can be calibrated for the \emph{GALEX} passbands into the
following form:
\begin{equation}
A_{FUV} = C(FUV-NUV)+ZP
\end{equation}
where FUV and NUV are apparent AB magnitudes in
  the respective \emph{GALEX} passbands \citep{calzetti01, seibert05,
  cortese06, goddard10}. For normal star-forming galaxies,
\citet{cortese06} give $C=5.12$, while \citet{seibert05} give a value
of $C=4.37$.  This value depends on the assumed star formation history
\citep[e.g.][]{calzetti05}, which affects the transformation from
$\beta$ to FUV$-$NUV color.  The value of $ZP$ depends on the age of
the regions of interest, and is relatively constant for populations
aged between $\sim$0--30 Myr \citep[Figure 9 in][]{goddard10}.
  We then derive the \ha \ extinction as A$_{\rm H\alpha} =
  0.5618$A$_{FUV}$, following Equation 13 of \citet{calzetti01}.

Following \citet{goddard10}, we make bulk radial extinction
corrections using the median FUV$-$NUV color of \ion{H}{2} regions
(hence excluding DIG and field O and B stars) in both M101 and
NGC~5474.  For ease of comparison, we adopt the same values of
$C=4.82$ and $ZP = 0.0$ as \citet{goddard10}, which are, respectively,
the average of the values of $C$ published in \citet{calzetti01,
  seibert05, cortese06}, and the typical color of $\sim$10 Myr old
populations \citep[Figure 9 in][]{goddard10}.  We find that our
results are not sensitive to these choices for reasonable values of
both.  The primary purpose of this correction is not to accurately
account for dust effects from \ion{H}{2} region to \ion{H}{2} region,
but rather to make a reasonable bulk correction that places the inner
and outer disks at the same mean extinction level for a more
consistent comparison among environments.  This is particularly
pertinent in our study, in which we measure the scatter in \hafuv
\ from environment to environment; because we are comparing
populations across large radial expanses (e.g. M101's inner vs. outer
disk), a strong gradient could increase the scatter in a given radial
range.

\begin{figure}
  \centering
  \includegraphics[scale=0.60]{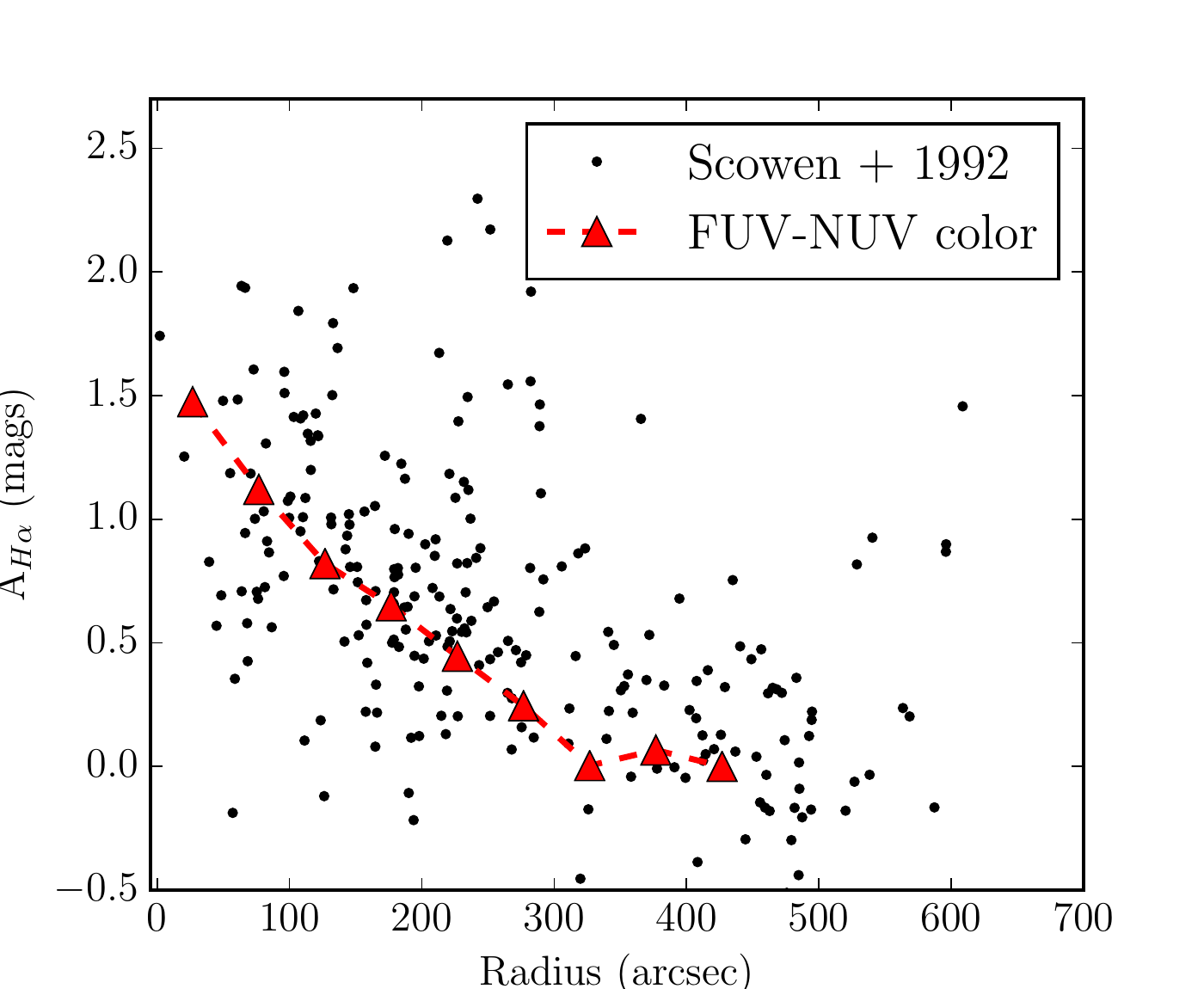}
  \caption[Extinction correction comparison]{H$\alpha$ extinction
    values derived from the \emph{GALEX} FUV-NUV color IRX-$\beta$ relation
    (red triangles), compared with the extinctions of \ion{H}{2}
    regions in M101 as derived from the Balmer decrement given by
    \citet{scowen92} (black points).
    \label{fig:scowen}}
\end{figure}

For comparison, we employed an alternative correction in M101 using
the extinction values published by \citet{scowen92}, derived from the
Balmer decrement.  We show this comparison in Figure \ref{fig:scowen}
by overplotting our UV color--derived values of $A_{H\alpha}$ to the
values for \ion{H}{2} regions from \citet{scowen92}, plotted as a
function of radius in M101.  While the two are broadly consistent, the
UV color--derived A$_{H\alpha}$ values are consistently lower by
$\sim$0.1 mag.  This is sensible, because the UV emission is directly
tracing the stellar populations, which may not always lie behind a
screen of dust depending on the relative dust geometry \citep[for a
  beautiful demonstration of this, see Figure 1 of][]{whitmore11}.
Because we are deriving the \ha \ extinction values by scaling
  A$_{FUV}$, this geometrical uncertainty also propagates into our
  values of A$_{\rm H\alpha}$.  However, we find through application
of both methods that this small offset does not affect the conclusions
of this paper.  We therefore use the UV color--derived values
throughout to maintain consistency.

\subsection{Region identification}
We use SEXtractor \citep{bertin96} to identify \ion{H}{2} regions
directly from the \ha \ difference mosaic.  Because we are selecting
regions based on their \ha \ emission, we are focusing our study only
on regions with ongoing star formation.  Our interest in this
particular study is in comparing physical differences in star
formation (for example, changes in the IMF) across environments, hence
by focusing on such short timescales, we avoid complications
introduced by aging populations, such as the dissolution of
H$\alpha$-emitting regions by stellar winds \citep{whitmore11}.

We perform photometry on all regions using a 4\farcs5 (150 pc)
  radius aperture, which is the typical FWHM of the \emph{GALEX} FUV
PSF (the Burrell Schmidt PSFs in the on- and off-band images have
FWHM$\sim$3\arcsec, hence the use of the FUV FWHM is warranted).  This
is large enough to contain multiple \ion{H}{2} regions at M101's
distance \citep[see, for example,][for sizes of Milky Way \ion{H}{2}
  regions]{quireza06}; we discuss how this affects our conclusions in
Section 4.3.  However, our statistical analyses are also robust to
moderate adjustments to the aperture size.  Additionally, we apply an
  aperture correction of 0.247 magnitudes to the FUV fluxes, derived
  from FUV-bright stars in the M101 field \citep[this agrees well with
    the curve of growth presented by][]{morrissey07}.

To efficiently pick out both outer-disk and inner-disk \ion{H}{2}
regions, we run SEXtractor at a 2$\sigma$ threshold on an
unsharp-masked version of our difference mosaic, without deblending.
This turns SEXtractor into something of a local peak-finding
algorithm, hence is useful for identifying the often densely packed
inner-disk \ion{H}{2} regions against the smooth background DIG.  That
said, it results in many spurious detections, thus we employ several
rejection criteria.  First, we run SEXtractor in dual-image mode,
measuring fluxes of difference image detections from the FUV images;
we reject all regions with $F_{FUV} \leq \sigma_{sky,FUV}$, where
$\sigma_{sky,FUV}$ is the pixel-to-pixel background dispersion in the
FUV images (measured from the intensity maps in the manner described
in Section 2.2).  We also reject any sources with \hafuv\ $>$ -1.2,
which is set by the maximum \hafuv \ value we find in Starburst99,
from a zero-age cluster with $1/50$ solar metallicity \citep[lower
  than the lowest metallicity found in M101;][]{croxall16}.  We also
reject sources with $g-r > 1.2$ and $r-i > 0.8$ to remove M stars
(Section 2.4).  Finally, we reject all sources $>$1440\arcsec \ (48
kpc) in radius from M101, and $>$360\arcsec \ (12 kpc) in radius from
NGC~5474.

These cuts remove the bulk of the contaminating sources.  However,
running SEXtractor with no deblending detects not only \ion{H}{2}
regions, but also local peaks in the DIG.  These regions are
identifiable by eye as being more uniform in flux across the
photometry aperture (as opposed to the point source--like \ion{H}{2}
regions).  However, to reduce subjectivity, we make a first-round
rejection of such regions via an automated procedure.  We define a
concentration parameter:
\begin{equation}
  c_{50} = 1-f_{{\rm px},50}
\end{equation}
where $f_{{\rm px},50}$ is the fraction of pixels in the photometry aperture
containing 50\% of the total flux ($c_{50}$ is defined such that high
values correspond to higher concentration).  We iterate the threshold
value of $c_{50}$ until we see a reasonable rejection of diffuse
regions, then reject the few remaining DIG regions by hand.  We choose
not to reject diffuse-looking regions in the outer disk; \ion{H}{2}
regions expand until they reach pressure equilibrium with the ISM
\citep{dyson80, garcia96}, hence in low density environments
they can potentially grow quite large.  The statistical analyses we
discuss below are robust to this rejection procedure, as
diffuse-looking regions most often have anomalously low \hafuv
\ \citep[which further implies they are mostly DIG;][]{hoopes01}, and
are rejected as outliers in the statistical metrics we use.

One concern is that in choosing regions based on \ha \ emission, there
is the possibility that we are missing a population of UV-bright but
\ha-weak clusters.  This would include, for example, very massive
clusters that nonetheless contain no highly ionizing, very massive
stars due to a truncated IMF.  We thus compared our \ha-selected
samples with separate samples selected from both galaxies' FUV images,
using the same procedure as before, however in this case, we rejected
regions based on their compactness in the FUV images rather than the
\ha \ image, in order to preserve FUV-emitting clusters.  While the
FUV-selected samples did uncover a large ($\sim$200) number of
additional very \emph{FUV-faint} regions (mostly in the outer disk),
as compared to the \ha-selected sample we found no significant number
of additional \emph{FUV-bright} regions at any radius.  All of the
additional FUV-faint regions selected also cover a wide range of \ha
\ flux (\lgflux $\sim$-14.6 -- -17 \fluxunit), and most appear diffuse
and irregular in the \ha \ difference image.  For example, many lie in
the diffuse outskirts of \ion{H}{2} region complexes, or along
filaments of more isolated diffuse emission.  This implies that they
could mostly be older FUV-emitting clusters embedded within the DIG.
It thus appears that if there is a population of FUV-bright but
\ha-faint regions in either M101 or NGC~5474, it is not significant
with respect to the general population of star-forming regions in
either galaxy at any radius.

\section{Results}

\begin{figure*}
  \centering
  \includegraphics[scale=0.60]{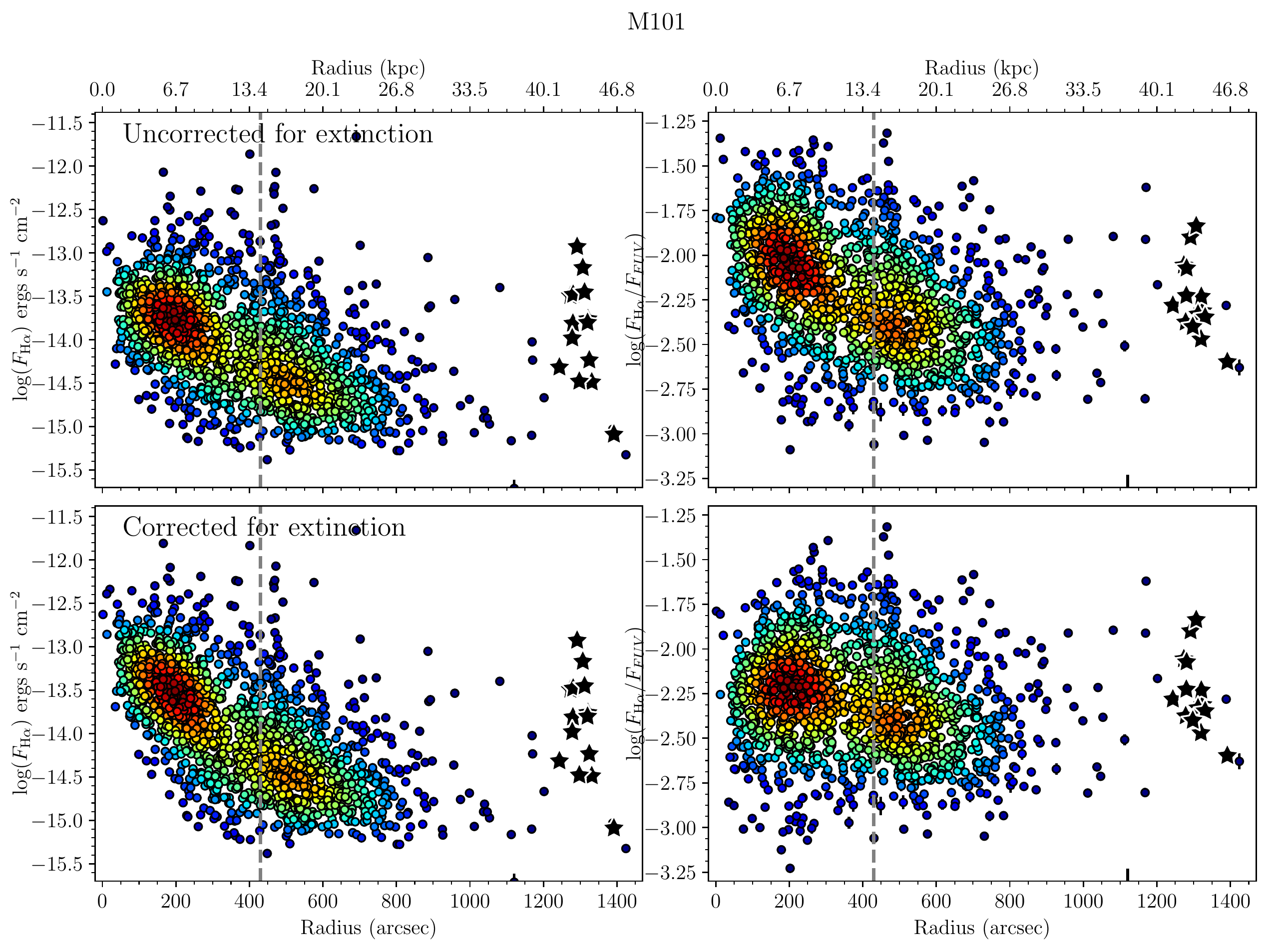}
  \caption[HII region fluxes and \hafuv \ ratios, M101]{Left:
    \ha \ fluxes of M101 \ion{H}{2} regions, plotted against radius.
    The top panels show fluxes uncorrected for extinction, while the
    bottom panels show fluxes after the correction described in
    Section 3.1 is applied.  The colors represent the local density of points
    in the plot.  Black stars represent regions located within the
    dwarf companion NGC~5477.  The gray dotted line shows our chosen inner
    disk--outer disk demarcation.  Right: \hafuv \ of all M101
    \ion{H}{2} regions, plotted against radius.  Symbol colors, symbol
    types, and the gray dotted line are the same as in the left plots.
    \label{fig:m101phot}}
\end{figure*}

\begin{figure*}
  \centering
  \includegraphics[scale=0.6]{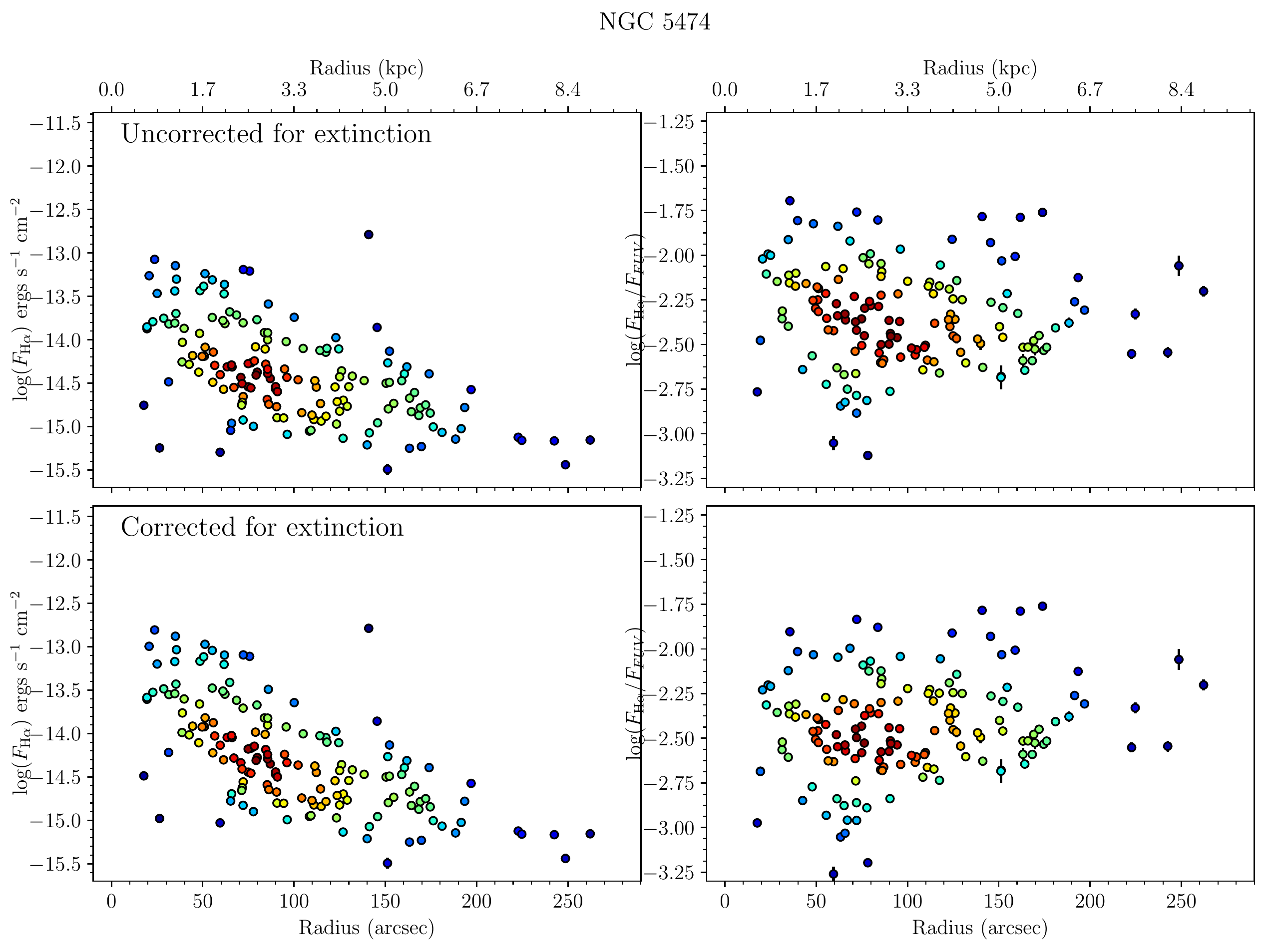}
  \caption[HII region fluxes and \hafuv \ ratios, NGC~5474]{As in
    Figure \ref{fig:m101phot}, but for \ion{H}{2} regions in NGC~5474.
    We have used the same scale on the y-axes for ease of comparison.
    \label{fig:n5474phot}}
\end{figure*}

\begin{figure*}
  \centering
  \includegraphics[scale=0.57]{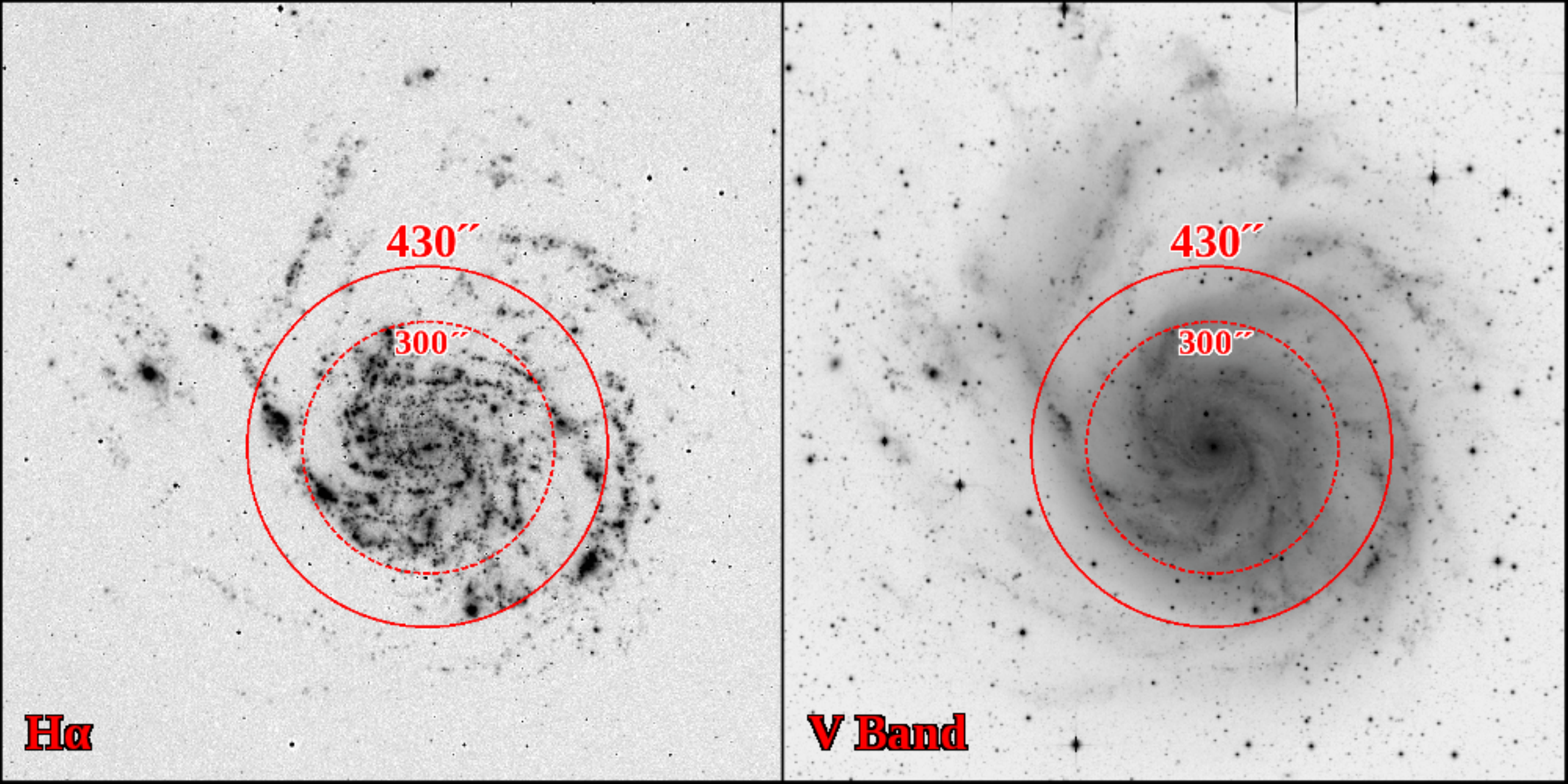}
  \caption[Inner disk/outer disk boundary in M101]{Choices of inner
    disk/outer disk boundary in M101, overlaid on the H$\alpha$
    difference image on the left, and the \emph{V}-band image
    \citep{mihos13} on the right.  The solid line marks our primary
    choice, which is 3 times the azimuthally averaged disk scale
    length (430\arcsec, or 14.5 kpc).  The dashed line marks an
    alternative (300\arcsec, or 10kpc), located where the H$\alpha$
    surface brightness profile begins to decline \citep{martin01}.
    \label{fig:rads}}
\end{figure*}

\subsection{\ion{H}{2} Region Photometry}

We show the results of our \ion{H}{2} region photometry in Figures
\ref{fig:m101phot} and \ref{fig:n5474phot} for M101 and NGC~5474,
respectively.  The final sample contains 1525 \ion{H}{2} regions in
M101 and 156 regions in NGC~5474.  We show radial profiles of \lgflux
\ on the left and $\log($\hafuv$)$ on the right.  For comparison, we
show radial profiles before and after we apply the extinction
correction described in Section 3.1 (top and bottom plots,
respectively).  The grey dashed lines in Figure \ref{fig:m101phot}
mark M101's outer disk, which we define as $>$3 times the azimuthally
averaged disk scale length \citep[430\arcsec, 14.5 kpc;][]{mihos13}.
In Figure \ref{fig:rads}, we show this outer disk demarcation and a
potential alternative on both our difference mosaic and on the
\emph{V}-band data from \citet{mihos13}, for reference.  We discuss
how the choice of outer disk boundary affects our results in Section
4.2.

It should be noted here that NGC~5474 has a strongly offset bulge
\citep{vanderhulst79, kornreich98}, hence the definition of its
``center'' is not entirely clear.  We define its center as the centroid
of the circular outer isophotes (at 180\arcsec, or 6 kpc) on our
on-band mosaic, which is very close to the kinematic center of its
(strangely regular) \ion{H}{1} velocity field \citep{vanderhulst79}.
This choice does not affect the qualitative behavior of the radial
profiles, however the flux profile does show more scatter with radius
when centered on the bulge.  This implies that the isophotal center is
the more appropriate choice regarding star formation in this galaxy.

The extinction correction has the expected behavior: \ha \ fluxes
increase absent extinction, and \hafuv \ decreases given stronger
attenuation for FUV.  The correction applied at all radii in the lower
metallicity companion NGC~5474 \citep[which has a central O
  abundance of $12 + \log({\rm O}/{\rm H})=8.19$, vs. 8.71 in
  M101;][]{pilyugin14} is less severe than that applied in the dustier
central regions of M101.  Additionally, we plot the values of \lgflux
\ and $\log($\hafuv$)$ for the dwarf irregular (dIrr) companion
NGC~5477 (due east of M101; see Figure \ref{fig:diffim}) in the same
plots as M101 using black stars.  Despite its much smaller mass,
NGC~5477's \ion{H}{2} regions span the same range of luminosity as
those in M101's inner disk, implying similar LFs between the two
environments.  The same appears true of NGC~5474; we found it possible
to reproduce NGC~5474's global LF by resampling from M101.  Each
galaxy contains pockets of high column density gas \citep[of order
  $10^{21}$ cm$^{-2}$;][]{vanderhulst79, vanderhulst01, walter08},
which may account for the similarity.  Regardless, that all three
galaxies have qualitatively similar LFs is reminiscent of the study by
\citet{schombert13}, which found that the lack of bright \ion{H}{2}
regions in LSB galaxies can be explained as an artifact of small
number statistics, rather than as a change in the LF itself.

Yet though each galaxy's integrated LF appears similar, there are
strong radial gradients in mean \ha \ luminosity in both M101 and
NGC~5474.  This is most likely a demonstration of the Schmidt Law:
molecular gas density in M101 declines exponentially with radius
\citep[e.g.][]{kenney91}, hence the SFR declines accordingly
\citep{kennicutt07, bigiel08}.  Also, the azimuthally-averaged SFR and
gas density within galaxies have a power law relationship \citep[up to
  the threshold density;][]{kennicutt98}, hence it is not surprising
that we see general radial declines in mean \ha \ flux with a large
region-to-region scatter.  A comparison with the THINGS \ion{H}{1} map
of M101 \citep{walter08} also shows that regions with the highest \ha
\ flux for their radius always cluster around high \ion{H}{1} column
density peaks.  That the global \ion{H}{2} region LFs of M101
and NGC~5474 (and possibly NGC~5477) appear similar thus seems a
consequence of each having a similar density structure within its
ISM.

If gas density alone imposes the radial dependence of \ha \ flux, it
should affect the FUV flux in a similar way, assuming no dramatic
changes in e.g. the IMF.  Indeed, Figure \ref{fig:m101phot} shows that
the radial gradient in \hafuv \ in M101 is strongly reduced after the
extinction correction is applied.  NGC~5474 contains no strong
gradient before correction; this remains mostly true after a
correction is applied, although a mild positive gradient is induced,
implying that perhaps we are slightly overcorrecting for extinction in
this galaxy.  We also find that \hafuv \ and F$_{FUV}$ are
  uncorrelated after applying an extinction correction in either
  galaxy.  Therefore, it may be that any radial trend in mean \hafuv
\ in either galaxy can be attributed to extinction.

The scatter in \hafuv \ also appears roughly constant with
environment, from M101's inner disk, to its outer disk, to NGC~5474,
and possibly even NGC~5477 (though with only 14 total \ion{H}{2}
regions, any measure of scatter in this galaxy will be highly
uncertain).  In tandem, this implies that star formation is ignorant
of the global environment; other than the available fuel, it does not
seem to know whether it is taking place in a low mass galaxy, a high
density inner disk, or a low density outer disk.  We test these
observations explicitly in the next section.

\subsection{Statistical Analysis}

\begin{figure}
  \centering
  \includegraphics[scale=0.65]{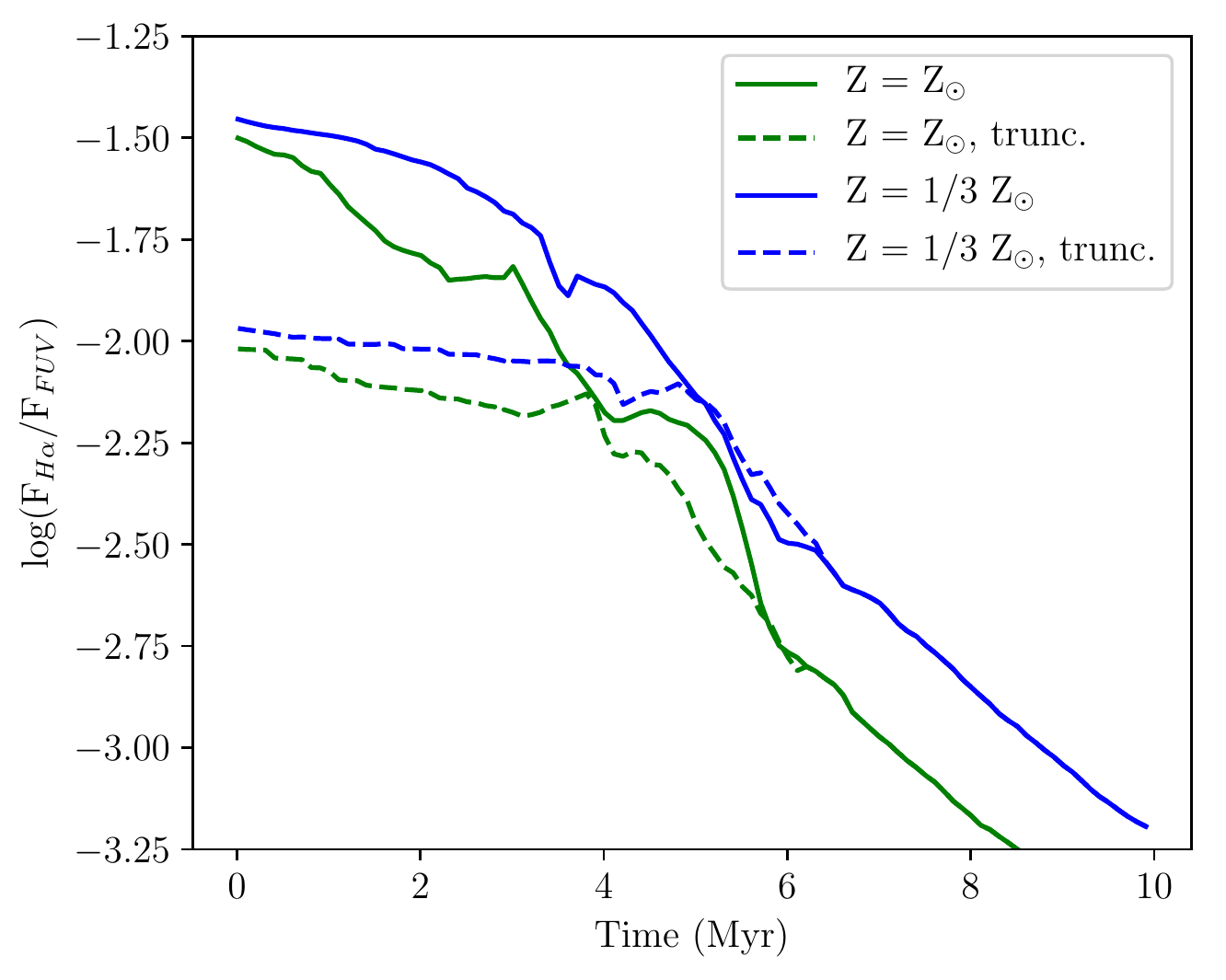}
  \caption[Evolution of \hafuv \ in Starburst99]{Starburst99
      single-burst models of the time evolution of \hafuv\ for two
    different metallicities \citep[using Padova
      isochrones;][]{bressan12}.  Solid lines show the evolution for a
    standard Kroupa IMF; dashed lines show a Kroupa IMF with a
    truncation at 30 M$_{\odot}$.
    \label{fig:sb99}}
\end{figure}

The intrinsic \hafuv \ ratio is mainly driven by the number of massive
O and B stars.  If present, they are the primary source of the
ionizing radiation that powers the H$\alpha$ emission.  A truncated
IMF would result in fewer massive stars being born, reducing the
maximum possible \hafuv.  We show this in Figure \ref{fig:sb99} via
evolutionary tracks of \hafuv \ in single-burst models from
Starburst99 \citep{leitherer99} for two different metallicities.  We
show both a standard Kroupa IMF \citep[solid lines;][]{kroupa01}, and
a Kroupa IMF truncated at 30 M$_{\odot}$ \citep[truncations as low as
  20 M$_{\odot}$ have been suggested, e.g.,][]{bruzzese15}.  The
tracks diverge clearly at early times ($\lesssim$6 Myr, set by the
lifespans of the most massive stars), with the truncated IMF tracks
peaking at much lower \hafuv \ as expected.  If a change in the IMF
occurs at a given radius in a galaxy, the distribution of allowed
values of \hafuv \ in the \ion{H}{2} region population will adjust
accordingly.  Lower variation in the region-to-region dust content in
outer disks would result in a similar change, once bulk radial trends
are taken into account.  Such behavior ought to be observable,
therefore, in the \emph{scatter} of bulk extinction-corrected \hafuv
\ within different populations of \ion{H}{2} regions, assuming that
variations in the median \hafuv \ can be fully attributed to
extinction effects.

\begin{figure}
  \centering
  \includegraphics[scale=0.65]{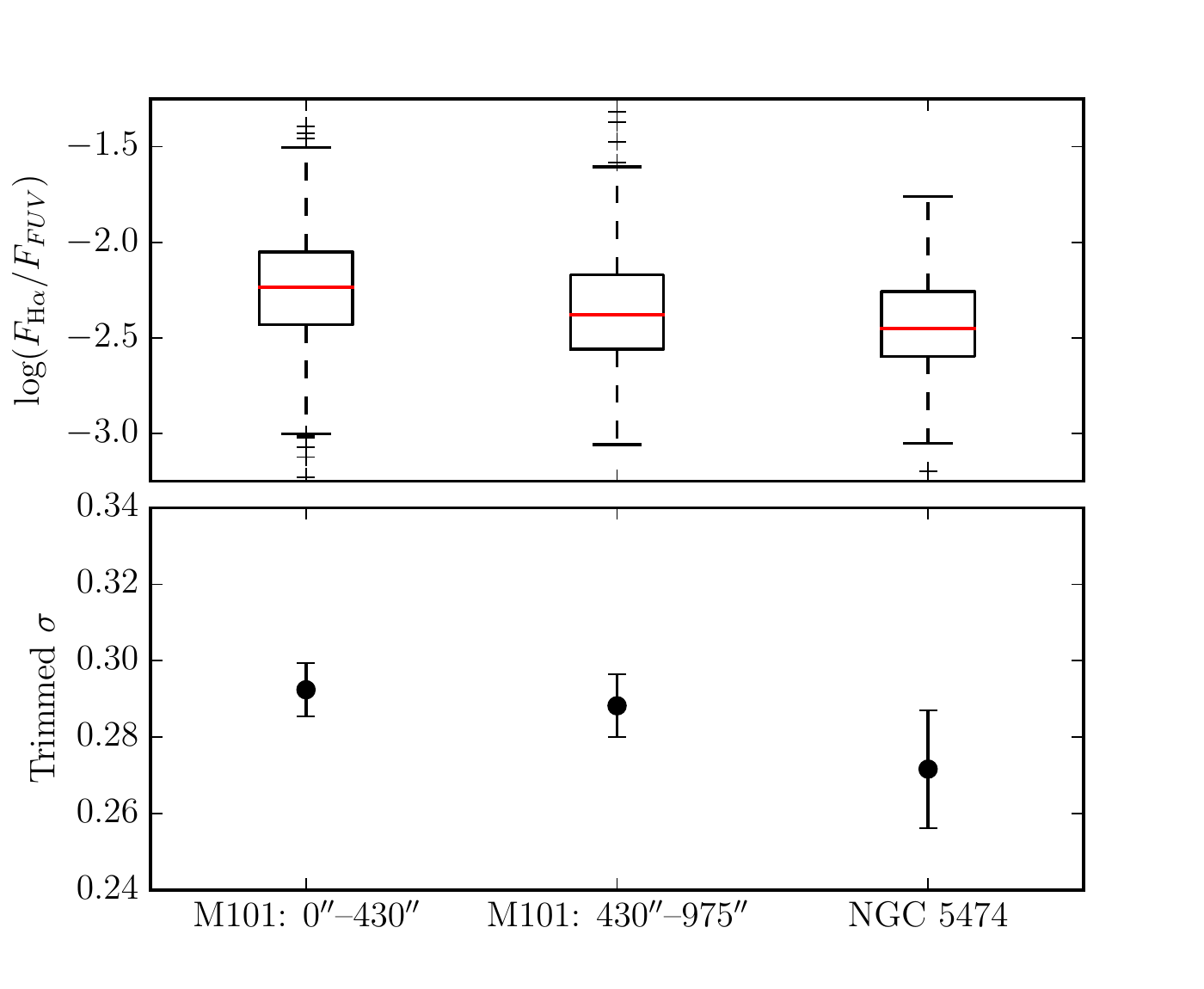}
  \caption[Statistical comparisons of \hafuv \ across the M101 Group]{Top:
    box and whisker plots showing the distribution of \hafuv \ values
    in three regions in the M101 Group: M101's inner disk, M101's
    outer disk, and NGC~5474 as a whole.  Bottom: values of the
    trimmed standard deviation in \hafuv \ for the three regions
    described above.  Error bars are $1/\sqrt{2N}$ for sample size
    $N$.  The radius 430\arcsec \ is 3$\times$ the scale length of
    M101.
    \label{fig:boxplots}}
\end{figure}

We display the medians and two measures of the scatter in \hafuv \ in
Figure \ref{fig:boxplots} for three regions: M101's
inner disk (inside of $3h$), M101's outer disk (outside of $3h$), and
the more massive companion NGC~5474.  The inner/outer disk
boundaries in M101 are marked on both Figures \ref{fig:m101phot} and
\ref{fig:rads}, for reference.

In the top panel of Figure \ref{fig:boxplots}, we show box and
whisker diagrams for these three regions.  As a reminder, the boxes
span the 1st through 3rd quartiles of the data (Q1 and Q3), and the
whiskers span to $\pm$1.5$\times$ the interquartile range.  Medians
are shown in red, and outliers as $+$'s.  In the bottom panel, we show
the values of the trimmed standard deviation ($\sigma_{\rm t}$) for the
same three regions.  This is the standard deviation of the sample
trimmed of its top and bottom 5\% of values, multiplied by a
corrective factor \citep[$1/0.789$ for 5\% trimming;][]{breiman73,
  huber81, morrison90} to ensure that $\sigma_{\rm t}$ and $\sigma$ (the
standard deviation of the whole sample) are measuring the same
parameter in the case of purely Gaussian data.  We use $\sigma_{\rm t}$
over $\sigma$ for its robustness to outliers, such as extremely
luminous \ion{H}{2} regions or the handful of DIG regions that might
have made it into the final sample; other such robust estimators of
scatter (such as the median absolute deviation) give similar results.
The error bars on $\sigma_{\rm t}$ are simply the standard error on the
standard deviation, which is equal to $1/\sqrt{2N}$ for sample size
$N$.

After we apply our extinction correction, the median values of \hafuv
\ for all three regions are: -2.235$\pm$0.013, -2.379$\pm$0.015, and
-2.451$\pm$0.029, respectively.  While this implies statistically
significant differences in the medians from region to region, we give
only the standard errors (which are equivalent to bootstrapped errors,
despite the slight non-Gaussianity of the data).  Systematic errors on
the \emph{GALEX}-calibrated IRX-$\beta$ extinction correction are
larger \citep[of order 0.1 mag, excluding uncertainties in the
  transformation from FUV$-$NUV color to $\beta$;][]{cortese06},
which does not include methodological uncertainty inherent in applying
this correction on average in radial bins.  The differences in the
medians between all three regions are also smaller than the standard
deviations in \hafuv \ ($\sigma \sim 0.3$), again implying that most
of the gradient in \hafuv \ in M101 can likely be explained by
extinction alone.

Similar box widths in Figure \ref{fig:boxplots}, as well as similar
values of $\sigma_{\rm t}$, also suggest that the scatter in \hafuv
\ among the three regions are equal.  We therefore compare the sample
variances using Levene's Test \citep{levene60}.  This test assesses
whether or not the quantity $z_{ij} = |x_{ij} - \bar{x_{i}}|$, where
$\bar{x_{i}}$ is the mean of the i-th group, is equal between groups.
It is hence similar to the F-test in that it assesses equality of
variances between populations, but it is more robust to
non-Gaussianity and higher in statistical power
\citep[e.g.][]{limloh96}.  The mean can be replaced with a more robust
statistic, such as the median \citep[e.g.][]{brown74}; we use the
trimmed mean, defined analogously to the trimmed standard deviation.

We show the results of this test in Table \ref{tab:levene} for the
following comparisons: M101's inner disk to its outer disk, M101's
inner disk to NGC~5474, M101's outer disk to NGC~5474, and all three
simultaneously.  $W$ is the value of the test statistic, while the
p-value is defined in the standard way for confidence $1-\alpha$.  In
all four tests, we cannot reject the null hypothesis that the variances
in \hafuv \ in all three environments are equal.  While, for
philosophical reasons, this does not by itself prove that the variances
are equal, these results in conjunction with the similarity in values
of $\sigma_{\rm t}$ and widths of the box plots for each region strongly
imply that this is the case.  We verified that this result is not sensitive
to the definition of the outer disk; the conclusion remains true for
choices anywhere between 300\arcsec \ \citep[the point at which the
  H$\alpha$ surface brightness profile begins to decline;][]{martin01}
and 600\arcsec \ \cite[roughly the Holmberg radius,
  R$_{26.5}$;][]{mihos13}.

\begin{deluxetable}{l c c c c}
\tablewidth{0pt}
\tabletypesize{\scriptsize}
\tablecaption{Results of Levene's Test Trials \label{tab:levene}}
\tablecolumns{2}
\tablehead{
  \colhead{TEST:} & \colhead{In-Out} & \colhead{In-5474} &
  \colhead{Out-5474} & \colhead{All}}
\startdata
\\
W & 0.293 & 0.009 & 0.052 & 0.147 \\[0.01cm]
p-value & 0.588 & 0.924 & 0.821 & 0.863
\enddata
\end{deluxetable}

While the comparisons between the regions of M101 and NGC~5474 seem
immune to the choice of inner disk--outer disk boundary, large
uncertainties in regions with smaller sample sizes could make it
harder to draw such a strong conclusion.  We thus further tested this
through a bootstrapping experiment.  For each definition of the inner
disk--outer disk boundary, we randomly sampled $N$ values of \hafuv
\ from either the inner or outer disk, with $N$ equal to NGC~5474's
sample size.  We then ran Levene's Test again between the
downsampled M101 population and NGC~5474.  We repeated each sampling
test 10000 times; in all tests, the resulting p-values were $>$0.05
between 93\% and 97\% of the time, providing evidence that the results
of the previous tests using the full samples were not an artifact of
sample size.

While these results are robust to the choice of inner
disk--outer disk boundary, we find that the lowest p-value was obtained
using 300\arcsec \ rather than 430\arcsec \ (p$=$0.14 vs. 0.59,
respectively).  By splitting the disk into three parts, we found that
the region within 300\arcsec--430\arcsec \ does have significantly
higher scatter in \hafuv.  Figure \ref{fig:m101phot} shows that this
region has a low density of \ion{H}{2} regions relative to the rest of
the disk.  It also appears dynamically distinct; it lies roughly at
co-rotation with the inner disk spiral arms \citep{waller97}, and is
the site of a severe kink in the \ion{H}{1} rotation curve
\citep{meidt09}.  This is also the location of a pocket of high
velocity gas in the galaxy's northeast \citep{walter08, mihos12} and a
region with a high velocity dispersion \citep{walter08}.  Dynamical
effects may thus have influenced the \ion{H}{2} region population in
this particular area \citep[a high gas velocity dispersion, for
  example, may inhibit star formation;][]{kennicutt89}.

Aside from this unusual region, however, we find that once extinction
is taken into account, both the median \hafuv \ and the scatter in
\hafuv \ shows no significant variation with environment in the M101
Group.  This supports our initial conjecture that, aside from gas
density (which affects the intensity of the star formation), star
formation on short timescales is blind to environment.

\subsection{Comparisons with Starburst99 Models}

\begin{figure*}
  \centering
  \includegraphics[scale=0.68]{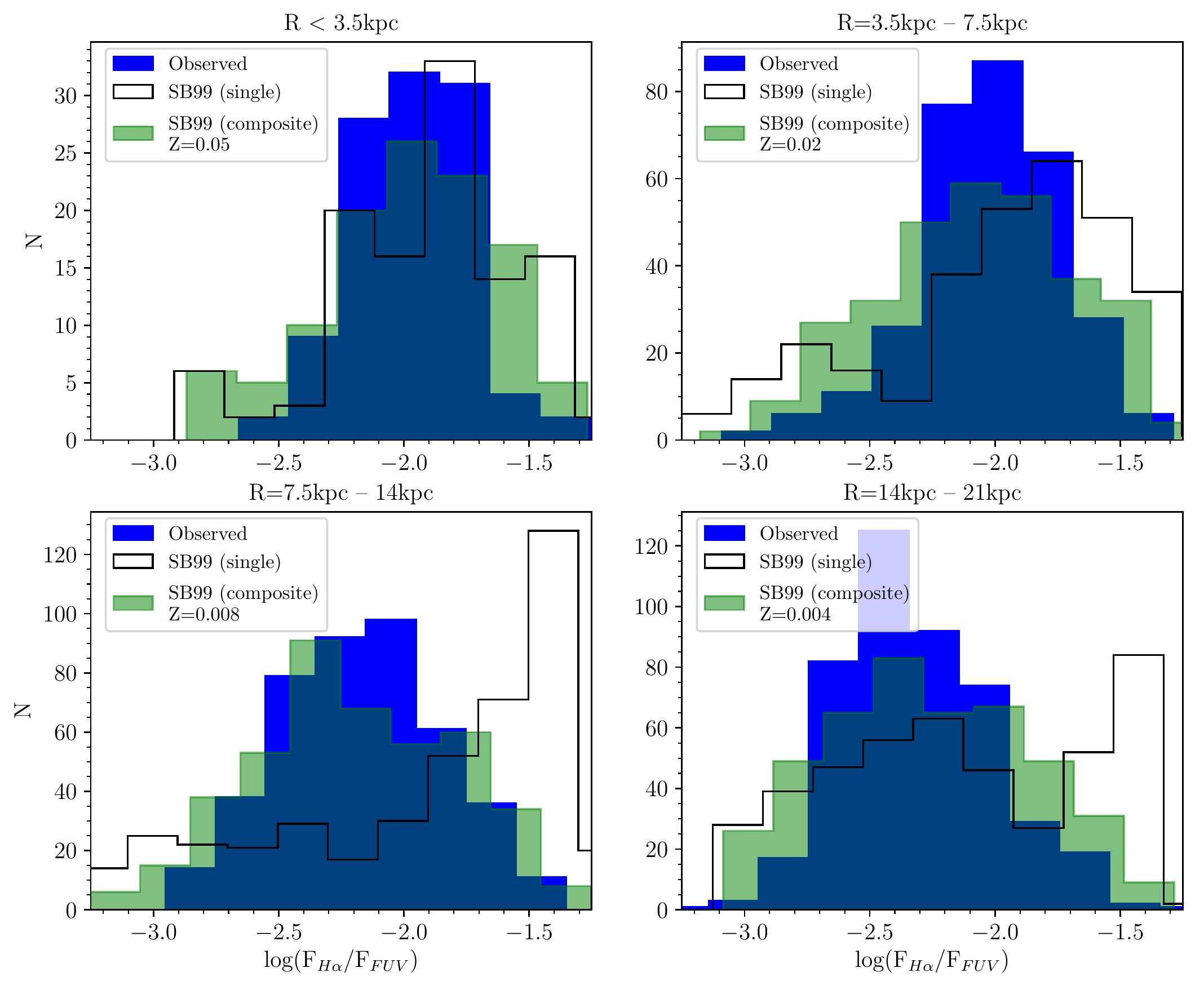}
  \caption[Observed and model distributions of \hafuv \ in
    M101]{Comparisons of the observed distributions of \ion{H}{2}
    region \hafuv \ (blue histograms) in M101 in different radial bins
    with model distributions from Starburst99.  Empty histograms show
    Starburst99 models with uniform sampling of single model regions,
    sampled from models with metallicities representative of their
    respective radial bins, while green histograms show averages of
    composite regions made of multiple Starburst99 models (see
      text).  \label{fig:sbhists}}
\end{figure*}

If both the median and scatter in the \hafuv \ ratio in \ion{H}{2}
region populations is constant with environment, once extinction is
taken into account, one might question how much room is left for
variations in the IMF.  We explore this question through comparisons
with Starburst99 \citep{leitherer99} models, which we show in Figure
\ref{fig:sbhists}.

\begin{figure*}
  \centering
  \includegraphics[scale=0.68]{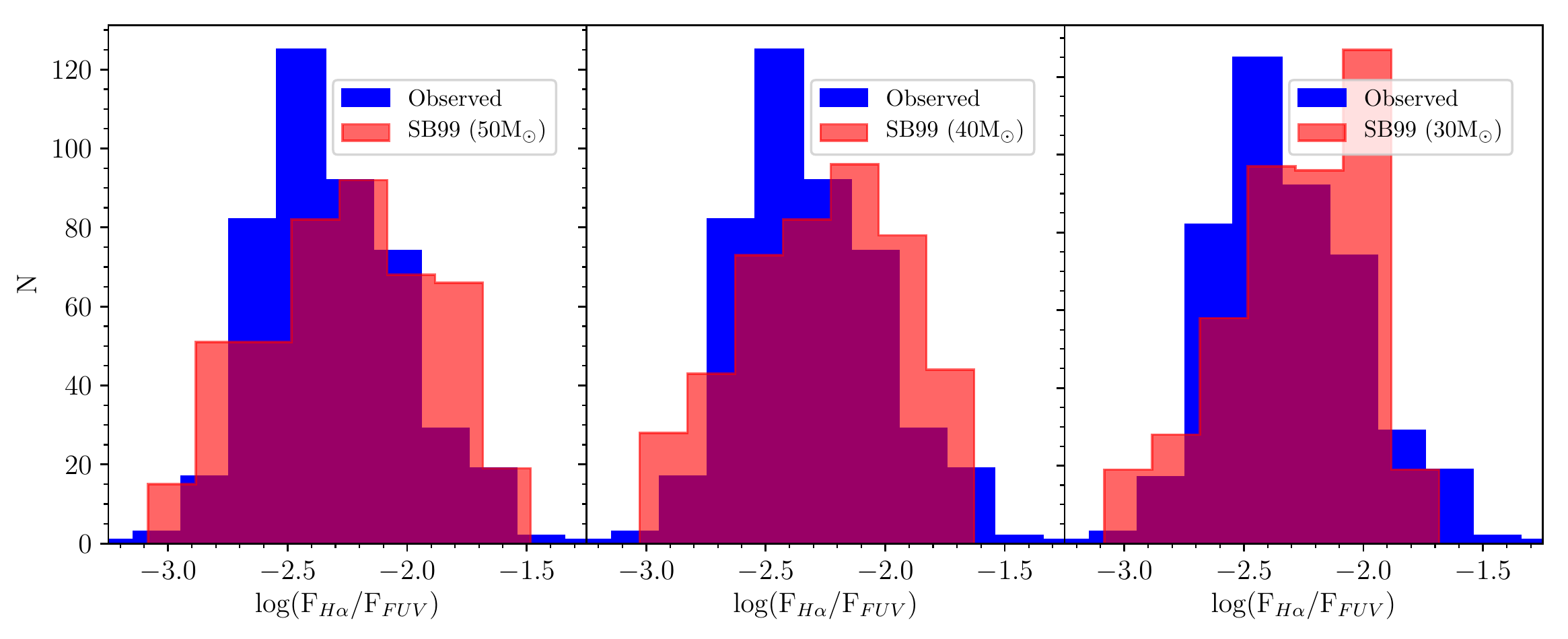}
  \caption{Comparisons of truncated Starburst99 models with the
      distribution of \hafuv \ found in the lowest metallicity
      (outermost radial) bin in M101 shown in Figure
      \ref{fig:sbhists}.  The real distribution is shown as blue
      histograms, which are the same in each panel.  Red histograms
      show models with mass truncations at 50 M$_{\odot}$ (left
      panel), 40 M$_{\odot}$ (middle panel), and 30 M$_{\odot}$ (right
      panel).  \label{fig:sb99trunc}}
\end{figure*}

Blue histograms in Figure \ref{fig:sbhists} show the distributions of
\hafuv \ in \ion{H}{2} regions in four radial bins within M101,
uncorrected for extinction.  In order to compare our data with
Starburst99 models, we chose these bins such that their mean
metallicities \citep[measured from the \ion{H}{2} region metallicity
  values supplied by][which range from $\sim$5$\times$ solar to $\sim
  1/5$ solar]{scowen92} corresponded to the metallicity options
available in Starburst99.  While Starburst99 does not include the
  effects of stochastic sampling from the IMF, given our low
  resolution the majority of the regions we sample are likely massive
  enough to be above the stochastic limit \citep[at M101, below
    $\log($F$_{\rm H\alpha}) \sim$ -15, e.g.][]{hermanowicz13}, hence
  these effects should not be important.

To generate model samples, we first create model evolutionary tracks
of \ha \ flux (using the output \ha \ luminosity) and FUV flux (by
convolving the output model spectrum at each timestep with the FUV
transmission curve) normalized to unit mass for each of the four
metallicities.  To create a realistic cluster sample, we create a
random distribution of masses following a power law with a slope of -2
(e.g. Lada \& Lada 2003; Hunter et al. 2003; Weidner et al. 2004; we
note that the results are robust for any reasonable choice of the
slope value), uniformly sample the model fluxes in time between 0 and
10 Myr, and multiply these mass-normalized fluxes by the randomly
generated cluster masses to produce a range of model cluster fluxes.
We then apply extinctions to these model fluxes at random, drawn from
the dataset by \citet{scowen92} within the appropriate radial bins,
and then trim the generated model regions to ensure that the
distribution of model \ha \ fluxes matches that of the data in each
radial bin.  We also reject model regions with FUV fluxes below the
observational limit.  The results are shown as the empty histograms in
Figure \ref{fig:sbhists}.

It can be easily seen in Figure \ref{fig:sbhists} that these
distributions provide a poor match to the data in all radial bins.  As
noted in Section 3.2, our choice of photometry aperture (4\farcs5)
corresponds to $\sim$150 pc at M101's distance, and is hence large
enough to potentially include multiple \ion{H}{2} regions (as well as
surrounding DIG).  We verified this through visual comparison with
archival HST \ha \ imaging of M101 (GO13773, PI Chandar), and found
that our apertures contain typically 4--5 individual \ion{H}{2}
regions.  Adjacent \ion{H}{2} regions should be similar in
metallicity, but may not be uniform in age; the Orion Nebula complex,
for example, contains four stellar associations within a $\sim$100 pc
radius that span ages from 0--10 Myr \citep{brown94}, arguing that any
individual \ion{H}{2} region complex identified in our sample may
actually consist of multiple clusters with varying ages.

Indeed, we found that we could reproduce the observed distributions
much more successfully by using model clusters generated by
  adding together $N$ individual Starburst99 models of varying ages,
  where $N$ is drawn from a Poissonian distribution with expectation
  $\lambda=4$.  These are shown via the green histograms in Figure
\ref{fig:sbhists}, where we have adopted a standard Kroupa IMF.
  While we cannot exactly reproduce the true distributions of \hafuv
  \ in any radial bin, this is perhaps not surprising given the large
  number of assumptions we have made (single metallicities per radial
  bin, randomly sampled extinction values, etc.).  Still, the median
  values of the model distributions are close to the true values
  (within ~0.1 dex in all radial bins), and the models tend to share
  the skewed Gaussian appearance of the true distributions.  Given the
  qualitative nature of these comparisons, however, we must address
  two caveats.

  First, we note that in order to reproduce the observed
    distributions in the two outermost radial bins, we must employ \ha
    \ and FUV flux cuts on both the low and high ends.  We find that
    if we trim only low fluxes, to match our observational limits, we
    cannot reproduce the distributions of \hafuv \ in the two
    outermost radial bins regardless of which IMF we choose.  Pure
    random sampling from the cluster mass function results in too many
    bright clusters for these outermost regions of M101.  This result
    is in agreement with \citet{pflamm13}.

    Second, even with these cuts the match appears poorest in the
    two outermost bins.  As has been previously argued, because outer
    disks seem to lack molecular gas and show extremely inefficient
    star formation, the IMF in such environments may be biased toward
    low mass stars \citep[e.g.][]{meurer09, pflamm09, bigiel10}.
  Hence we attempted to determine whether or not a truncated IMF
  provided a better match to this bin.  This is shown in Figure
    \ref{fig:sb99trunc}, where we compare the distribution of \hafuv
    \ values in the lowest metallicity (outermost) radial bin to three
    different models, truncated at 50 M$_{\odot}$ (left panel), 40
    M$_{\odot}$ (middle panel), and 30 M$_{\odot}$ (right panel).
    Again, these models are averages of typically 4-5 model regions.
    We find that 50M$_{\odot}$ is the lowest truncation mass we can
    use in order to produce satisfactory qualitative agreement with
    the data.  Below this mass, the model distributions of \hafuv
    \ tend to be skewed strongly to the left and consistently lack
    high \hafuv \ tails.  The lower the truncation mass, the sharper
    the cutoff at high \hafuv.

  Figure \ref{fig:sb99} provides an explanation: the time evolution of
  \hafuv \ for truncated IMFs shows a plateau at early ages, the
  length of which depends on the lifetime of the highest mass star,
  beyond which \hafuv \ begins to decline.  The plateau value itself
  also depends on the mass, such that lower truncation masses plateau
  at lower values of \hafuv.  The standard Kroupa IMF model, by
  contrast, shows a steady decline over a larger range of \hafuv
  \ values; the decay in \hafuv \ reflects the larger range of
  contributing stellar masses, hence the larger range of stellar
  lifespans.  Uniform sampling in time from the truncated
  distributions thus results in a distribution of \hafuv \ that is
  strongly peaked at the plateau value. Extinction adjusts the model
  values of \hafuv \ slightly higher, but is not strong enough in the
  outermost bin to create a noticeable high-\hafuv \ tail.

In summary, we find that while IMFs truncated as low as 50 M$_{\odot}$
can qualitatively reproduce the observed distributions of \hafuv \ in
\ion{H}{2} regions throughout M101, they produce no better
  agreement than a standard Kroupa IMF.

\section{Discussion}

We have shown that the distribution of the \hafuv \ ratio in
\ion{H}{2} region populations, aside from extinction effects, does not
change with environment in the M101 Group.  We have also shown that we
can model the observed distributions of \hafuv \ in \ion{H}{2} regions
throughout the M101 Group without invoking a truncated IMF.  The
makeup of stellar populations ionizing \ion{H}{2} regions throughout
the M101 Group therefore appears ignorant of the local surface mass
density; only the intensity of star formation changes.

We thus consider the possible origin of trends in \hafuv \ with
e.g. galaxy stellar mass and central surface brightness found by other
authors \citep[e.g.][]{lee09, meurer09}.  Because these studies focus
on the integrated \ha \ and FUV fluxes of galaxies---which includes
compact \ion{H}{2} regions, DIG, and diffuse FUV emission---we
consider how the DIG, diffuse FUV emission, and bias in measurement
techniques might each contribute to the observed trends in integrated
\hafuv \ of whole galaxies.  Finally, we discuss our results in the
context of the M101 Group itself, particularly its tidal interaction
history, and consider whether or not our results for this group can be
generalized to other systems.

\subsection{On the Observed Trends of Integrated \hafuv}

If the IMF does not change with environment, as we have argued, why
then do many studies find that \hafuv \ integrated over galaxies, or
azimuthally averaged in wide radial bins, is lower in low density
environments \citep[e.g.][]{gildepaz05, thilker05, lee09, meurer09,
  goddard10}?  Because the integrated \hafuv \ includes all sources of
\ha \ and FUV emission, from \ion{H}{2} regions to DIG to diffuse FUV
with no \ha \ counterpart, changes in integrated \hafuv \ can result
from many different factors, from variations in the IMF, to stochastic
sampling of the IMF in low mass clusters, to a non-uniform SFH
\citep[e.g.][]{lee09, pflamm09, alberts11, barnes11, barnes13,
  weisz12, dasilva14}.

We have shown that variation in the IMF is unlikely within the M101
Group.  Given our low resolution, most ($>$90\%) of the \ion{H}{2}
region complexes we identified have fluxes above where stochastic
sampling ought to be important \citep[e.g.][]{hermanowicz13}.
However, a non-uniform SFH could could result in abundant
populations of FUV-emitting stars with no \ha \ counterpart,
specifically if such populations are remnants of a fading burst of
star formation.  Therefore, given that \hafuv \ is roughly constant in
\ion{H}{2} regions, we test for an overabundance of FUV
relative to \ha \ by comparing the fractions of \emph{diffuse} \ha \ emission
(or the DIG fraction, hereafter $f_{\rm DIG}$) and \emph{diffuse} FUV
emission (hereafter, $f_{\rm DUV}$).  We define ``diffuse'' emission
as any \ha \ or FUV emission located outside of what we have defined
as \ion{H}{2} regions, for simplicity; hence, we measure $f_{\rm DIG}$
and $f_{\rm DUV}$ by masking out \ion{H}{2} regions.

For the purposes of this study, we are concerned mainly with the
relative values of the diffuse fractions between environments, hence
it is important only that we measure $f_{\rm DIG}$ and $f_{\rm DUV}$
in a consistent manner for each environment.  However, given that we
base our mask on the low-resolution GALEX imaging, it is useful to
compare our value of $f_{\rm DIG}$ with that found in other studies
to estimate how much of the DIG directly adjacent to \ion{H}{2}
regions we could be masking.  Given our canonical mask (4\farcs5
\ apertures), we find a DIG fraction of 33\%.  Our masking thus
appears more aggressive than previous studies of the DIG
\citep[e.g.][who found a DIG fraction of 43\% in M101]{thilker02},
implying that with our canonical mask we are isolating the most
diffuse part of the DIG.  Changing the mask aperture by $\pm$1\farcs5
\ yields changes in $f_{\rm DIG}$ and of $f_{\rm DUV}$ of $\pm$20\%.

To test the influence of this masking on the relative fractions
of DIG and diffuse FUV , we measured the azimuthally averaged radial
profile of \hafuv \ in the DIG in both M101 and NGC~5474.  We found
that both galaxies display a distinct downward trend in \hafuv \ with
radius---implying dominant diffuse FUV emission in their outer
disks---which persists even when using an unrealistically aggressive
mask that results in $f_{\rm DIG} \sim $6\% for M101.  Therefore, our
masking procedure does not appear to influence the results we present
here.  We defer a more detailed discussion of the DIG in the M101
Group to a forthcoming paper.

While the expectation is that high $f_{\rm DUV}$ relative
to $f_{\rm DIG}$ should yield lower integrated \hafuv \ ratios, we
find that this is not always the case, implying that there may be
methodological bias at play as well in measuring integrated properties
of galaxies and regions of galaxies.  Specifically, a bias may be
incurred when using flux-weighted values over, e.g., areal-weighted
values of \ha \ and FUV flux.

\begin{deluxetable*}{l c c c c c}
\tablewidth{0pt}
\tabletypesize{\scriptsize}
\tablecaption{Integrated Properties of M101 Group Galaxies \label{tab:int}}
\tablecolumns{6}
\tablehead{
  \colhead{Region:} & \colhead{M101} & \colhead{Inner M101} & 
  \colhead{Outer M101} & \colhead{NGC~5474} & \colhead{NGC~5477}
}
\startdata
\\
$f_{\rm DIG}$ & 0.33 & 0.34 & 0.28 & 0.17 & 0.20 \\[0.01cm]
$f_{\rm DUV}$ & 0.56 & 0.51 & 0.60 & 0.42 & 0.45 \\[0.01cm]
\hafuv & -1.98 & -2.08 & -1.95 & -2.29 & -2.14  \\[0.01cm]
\hafuv, corr. & -2.21 & -2.22 & -2.21 & -2.48 & -2.22 

\tablecomments{
  Rows are: fraction of \ha \ flux from the DIG (1), fraction of
  diffuse FUV emission (2), integrated \hafuv \ (3), and integrated
  \hafuv \ corrected for extinction (4).  Systematic uncertainties,
  which dominate, are discussed in the text.
  }
\enddata
\end{deluxetable*}

In Table \ref{tab:int}, we give diffuse fractions in five
environments in the M101 Group: M101 as a whole, M101's inner disk,
M101's outer disk, its more distant companion NGC~5474, and its
nearby dIrr companion NGC~5477.  We measure both $f_{\rm DIG}$ and
$f_{\rm DUV}$ in an identical manner, hence they are comparable
regardless of uncertainty in e.g. the choice of \ion{H}{2} region
mask.  Additionally, for each region we give integrated values of
\hafuv \ before and after applying an extinction correction.  In this
case, we apply an integrated correction, measured using the integrated
FUV$-$NUV colors of each region, as is typically done in galaxy survey
studies \citep[e.g.,][]{lee09, meurer09}.

Comparison of the diffuse fractions in \ha \ and FUV indicate that
diffuse FUV emission is more prevalent compared to DIG in M101's outer
disk and in both companion galaxies.  This concurs with a visual
examination of the images; in M101's outer disk, for example, we find
many large (several kiloparsec wide) patches of diffuse FUV emission
that have no \ha \ counterpart in our difference image.  The areal
covering fraction of diffuse FUV emission appears larger than the DIG
covering fraction across the whole outer disk, while in the inner
disk, the covering fractions of both are roughly equal.
Quantitatively, this is observable as a larger outer disk scalelength
in the FUV compared to the \ha, such as is typically seen in other XUV
disks \citep{gildepaz05, thilker05, goddard10}.

However, the integrated values of \hafuv \ do not reflect this.
Despite the larger fractions of older FUV-emitting populations in
M101's outer disk and in the two companion galaxies, after correcting for
extinction, only NGC~5474 shows a significantly different value of
integrated \hafuv.  This appears to be an artifact of the
flux-weighted measurement; in NGC~5474, we find that the brightest
10\% of \ion{H}{2} regions (only 16 regions) contribute nearly 60\% of
the galaxy's total \ha \ flux.  Thus, if something is systematically
different about these few regions---age, dust content---compared to
the remaining \ion{H}{2} regions in the galaxy, this difference will
drive the galaxy's flux-weighted mean \hafuv \ ratio to an
unrepresentative value.  In NGC~5474, the brightest \ion{H}{2} regions
have redder FUV$-$NUV colors ($\sim$0.05 compared to $\sim -$0.1 in the
dimmer regions).  Because we derive the extinction based on the UV
color, these regions are measured as dustier environments; if so, the
extinction correction may be overcompensating for dust throughout
NGC~5474 and driving the integrated \hafuv \ down.

This is demonstrated in an alternative way in M101's outer disk.  As in
NGC~5474, the brightest regions in M101's outer disk are redder in UV
color ($\sim -$0.05 compared to $\sim -$0.2), hence potentially dustier,
and again contribute a large fraction of the region's total \ha \ flux (40\%).
Before applying an extinction correction, the median \hafuv \ value of
all of the \ion{H}{2} regions in M101's outer disk is -2.32, but the
flux-weighted mean value of \ion{H}{2} regions is -2.09.
Flux-weighting thus drives the integrated \hafuv \ ratio of \ion{H}{2}
regions in M101's outer disk to a higher value, as it is biased by the
brighter, redder regions, in direct analogy with the integrated \hafuv
\ value of NGC~5474.

As such, it is unclear whether or not the trends in integrated \hafuv
\ with stellar mass, SFR, and surface brightness noted in other
studies result from physical changes or purely from systematics
induced by the flux-weighted measurements.  Regardless, the M101 Group
is a well-studied system, with constraints on stellar populations
throughout its disk and constraints on its tidal interaction history
with its companions \citep[e.g.][]{beale69, rownd94, waller97,
  mihos13}.  We can therefore make more specific conclusions about
how M101's local environment may have influenced the star formation
taking place in its outer disk and companions, and consider whether or
not these conclusions can be generalized to other similar systems.  We
discuss this further in the following section.

\subsection{The M101 Group As a Case Study}

We have shown that in the M101 Group, \ion{H}{2} regions have roughly
constant \hafuv \ distributions regardless of their environment.  We
have also shown that diffuse FUV emission, with no \ha \ counterpart,
is abundant in M101's outer disk and its two companions, implying
widespread populations of slightly older O and B stars in the field,
similar to other XUV disks \citep[e.g.][]{gildepaz05, thilker05}.  We
argue here that this can be explained in the context of M101's
interaction history, and consider whether or not star formation in the
low density environments of the M101 Group could be representative of
low density environments as a whole.

In general, the origin of field O and B stars is not yet clear.  They
may form in-situ \citep{dewit05, lamb10, oey13}, form within
\ion{H}{2} regions but be ejected at high velocity \citep{gies87,
  moffat98, dewit05}, or they may be young clusters that have fully
succeeded in clearing out gas and dust from their birth \ion{H}{2}
regions.  In a study of diffuse FUV emission in the interarm regions
of M101's inner disk, \citet{crocker15} found that the majority
is likely emitted by 10--50 Myr old stellar populations that have
drifted from their birthplaces in spiral arms.  Because these stars
are carried by the disk's underlying rotation, the difference
between the rotation speed and the spiral arm pattern speed determines
how far they might travel from a given spiral arm; one might expect
stars to remain very close to spiral arms near corotation, for
instance.

UV light scattered into our line of sight by dust contributes a
sizeable fraction of the diffuse UV as well \citep[upward of
  $\sim$60\%;][]{crocker15}, but only in the vicinity of spiral arms;
in a field adjacent to a spiral arm, \citet{crocker15} estimate that
the UV flux contributed by scattered light drops by a factor of
roughly 1.5 over a distance of $\sim$1.5 kpc.  This, along with lower
dust content, implies that diffuse FUV in M101's outer disk contains
very little scattered light.  For example, in the galaxy's northeast,
we find large patches (several kpc on a side) of diffuse FUV located
some 5--10 kpc from the nearest spiral arm, and some a similar
distance from the nearest \ion{H}{2} region.  This FUV emission thus
appears to be a remnant of a previous episode of star formation, which
either formed in-situ or migrated from elsewhere in the disk.  The
largest such patch ($\sim$2 kpc in radius, detected at $> 10\sigma$
significance in the FUV) has an FUV$-$NUV color of $\sim$0.6; in a
model of color evolution in integrated populations by
\citet{boissier08}, young populations maintain an FUV$-$NUV color of
$\sim$0.0 while SF is ongoing, and reach $\sim$0.6 roughly 200 Myr
after star formation begins to decline (neglecting extinction,
although extinction may be safely neglected in outer disks).  In the
Milky Way, populations of O and B stars have radial velocity
dispersions of order $\sim$10 km s$^{-1}$ \citep{binney98}, thus can
easily disperse over $\sim$2 kpc in radius in 200 Myr.  This diffuse
FUV-emitting starlight thus likely formed in a localized burst a few
hundred Myr ago and is now beginning to fade.  We find many other such
patches of diffuse FUV throughout M101's outer disk with similarly red
colors ($\sim$0.4--0.6), implying similar origins.

M101's disturbed morphology implies that it suffered a recent tidal
interaction.  From integrated $B-V$ colors in its outer disk,
\citet{mihos13} proposed that this morphology resulted from a fly-by
encounter with its more distant companion NGC~5474 some $\sim$300 Myr
ago, resulting in a brief and currently fading burst of star
formation.  After 300 Myr, even the NUV light begins to fade; \ha
\ emission would thus be scarce, as it is in the diffuse FUV patches
discussed above.  Follow-up HST imaging of stellar populations in
M101's northeast plume region are consistent with this star formation
timeline (Mihos et al. in prep.), providing strong support that star
formation in M101's outer disk was induced by an interaction.  This in
turn shows that M101's outer disk does not have a uniform SFH.  If
NGC~5474 was the culprit in the interaction, it too should have seen a
starburst on the same timescale, hence it too should have a
non-uniform SFH.  The M101 Group hence provides a fairly clear example
of an FUV-dominated outer disk resulting from a fading, tidally
induced starburst; from this perspective, too, it is not necessary to
invoke changes in the IMF to explain the star-forming properties of
the M101 Group.

Is this scenario generalizable to other systems?  XUV disks are often
suggested to be tidal in origin \citep{gildepaz05, thilker05,
  thilker07}, or else are created through gas accretion into the
outer disk \citep{lemonias11}.  Also, the UV emission in XUV disks is
typically concentrated in filments reminiscent of spiral structure
\citep{thilker07}.  While outer disks may typically be stable against
spiral arm formation, we can consider the longevity of a set of spiral
arms induced in an outer disk by a tidal interaction, hence the
longevity of XUV disks in general.  As a rough estimate, let us assume
that spiral arms in outer disks are not self-sustaining due to high
disk stability \citep[e.g.][]{kennicutt89} and so lose their coherency
over one dynamical time; in M101 at 16 kpc (roughly where we demarcate
its outer disk), this is $\sim$500 Myr \citep[assuming a rotation
  speed of $\sim$190 km s$^{-1}$;][]{meidt09}.  Star formation
persists in M101 out to $\sim$40 kpc, where a dynamical time is
$\sim$1.3 Gyr.  As such, if galaxies like M101 suffer only one
interaction in their lifetimes capable of producing an XUV disk,
assuming a total lifetime of $\sim$10 Gyr, there would be a
$\sim$5--13\% probability that we would witness it in this state at
$z=0$.  A study by \citet{lemonias11} found that XUV disks exist in
4--14\% of galaxies out to $z=0.05$; if M101 can be considered
representative \citep[it is slightly brighter than L$^{*}$ in the
  \emph{V}-band;][]{devau91}, then the average galaxy need suffer only
1--2 interactions capable of producing XUV disks in their lifetimes to
explain the frequency of XUV disks in the local universe.  Whether or
not this is reasonable depends on how specific the parameters of the
interactions must be in order to produce an XUV disk (mass ratio,
relative inclination, relative velocities, etc.), but this simple
argument suggests that all XUV disks may be explainable through tidal
interactions with satellites.

Even in purely isolated systems, the global stability of outer
disks requires that some manner of perturbation is still necessary to
initiate star formation there \citep[e.g. substructure in the dark
  matter halo;][]{bush10}.  If star formation in dwarf galaxies
results mainly from e.g. supernova feedback \citep{vanzee97}, some
manner of perturbation would be required to initiate it in the first
place.  This general dependence on external forces, rather than on
potentially long-lived, regularly rotating spiral features or bars,
implies that star formation in low density environments may always be
subject to stochasticity, hence an assumption of constant star
formation over Gyr timescales in such environments could be highly
suspect.  As more and more systems are studied, and as finer and finer
resolution SFHs are obtained of these systems, the nature of star
formation and the evolution of galaxies should begin to become clear.

\section{Summary}

We have performed a study of star formation across all environments in
the nearby M101 Group---M101's inner disk, its outer disk, and its two
lower mass companions---using both new deep \ha \ narrow-band imaging and
archival UV data \citep[\emph{GALEX} NGS and PI data;][]{bianchi03,
  bigiel10}, in order to test whether or not star formation physics
(specifically the IMF) changes with environment.  We have chosen to
study only the \ha-emitting \ion{H}{2} regions in these environments,
in order to compare only populations young enough to retain their most
massive stars.

We have performed photometry on all \ion{H}{2} regions in M101,
NGC~5474, and NGC~5477 in order to measure their \hafuv \ ratios,
which should be systematically low in the absence of massive ($M
\sim$20--100 $M_{\astrosun}$) stars.  We find that the median \hafuv
\ ratio across all populations of \ion{H}{2} regions in the M101 Group
does not vary signficantly, once bulk radial extinction trends are
taken into account.  In addition to the median, however, we also find
that the \emph{scatter} in \hafuv \ does not vary significantly with
environment.  While typical \ion{H}{2} region fluxes do decline with
radius in M101 and its larger companion NGC~5474, their near constant
distribution of \hafuv \ ratios implies that the populations of
ionizing stars even in the fainter outer disk \ion{H}{2} regions are
being sampled from the same IMF as in the inner disk.  The decline in
mean \ion{H}{2} region flux may thus be attributable primarily to a
decline in mean surface gas density alone, rather than any significant
change in the cloud-to-cloud physics of star formation.

Using Starburst99 models \citep{leitherer99}, we attempted to
determine whether or not a truncated IMF was necessary to reproduce
the observed distributions of \ion{H}{2} regions in any radial bin in
M101.  We find that, while we are able to qualitatively reproduce
these distributions using IMFs truncated at the high mass end
  ($\lesssim$50 $M_{\astrosun}$), we are able to just as successfully
reproduce the distributions of \hafuv \ using a standard Kroupa IMF,
regardless of the local surface brightness.  It therefore appears
that, at least when comparing bulk populations, it is unnecessary to
invoke changes to the IMF to explain the properties of the \ion{H}{2}
regions in M101.

Assuming the IMF is universal, we further investigate the origin of
trends in \emph{integrated} \hafuv \ with surface brightness, SFR, and
stellar mass found by other authors \citep[e.g.][]{thilker05,
  boselli09, lee09, meurer09, goddard10}.  Because the \hafuv \ ratios
of \ion{H}{2} regions do not change throughout the M101 Group, we
compare the relative fractions of \emph{diffuse} \ha \ and FUV
emission, and find that indeed, diffuse FUV emission with no \ha
\ counterpart---hence no extremely young, massive stars---is more
prevalent in lower surface density regions.  This implies that such
regions suffered a recent but now fading burst of star formation,
hence are not necessarily continuously forming new stars; low
integrated \hafuv \ ratios in these low density regions may thus
result from a bursty or otherwise non-uniform SFH.  However, we have
also shown that using flux-weighted \hafuv \ or flux-weighted
extinction corrections can bias the value of the integrated \hafuv
\ in galaxies, particularly if the bulk of the \ha \ or FUV intensity
emerges from a small number of bright \ion{H}{2} regions.  We thus
advise caution in future such studies with regard to how integrated
\hafuv \ is measured and corrected for extinction.

Finally, we consider whether or not the M101 Group could be
exceptional or whether our results are more broadly applicable.
Previous studies have shown that star formation in M101's outer disk
was likely triggered by a tidal interaction several hundred Myr ago
\citep{mihos13}.  This lends credence to the interpretation that the
abundant populations of FUV-emitting stars with no \ha \ counterpart
are remnants of a now fading burst of star formation.  Additionally,
we have shown that if interaction-induced star formation in outer
disks persists over only one dynamical time, it may still be
long-lived enough to account for the low frequency
\citep[4--14\%;][]{lemonias11} of XUV disks observed in the local
universe.  Therefore, assuming the conditions necessary to create XUV
disks through tidal interactions are not oddly specific, it is not
unreasonable to consider that all XUV disks may be tidally induced.
If so, this would imply that star formation in low density
environments only differs from star formation in high density
environments in that it requires outside perturbation to be initiated.
This implies in turn that star formation at low density is subject to
greater stochasticity than star formation at high density; assuming a
uniform SFH in the low density regime may be unwise.

\section{Acknowledgements}

This work has been supported by the National Science Foundation,
through award 1108964 to J.C.M., as well as through award 1211144 to
P.H.  This work made use of Numpy, Scipy \citep{oliphant07},
MatPlotLib \citep{hunter07}, astroML \citep{vanderplas12}, and the
community-developed core Python package for astronomy, Astropy
\citep{astropy13}.  We would like to thank Heather Morrison for
many helpful discussions regarding the statistical analyses presented
in this work, as well as Stacy McGaugh, Sally Oey, and Daniella
Calzetti for helpful discussions about extinction, star formation
physics, and star formation tracers.  We would also like the
  thank the anonymous referee for a very thorough and helpful review,
  which we feel greatly improved this manuscript.

{\it Facilities:}
\facility{CWRU:Schmidt} - The Burrell Schmidt of the Warner and Swasey
Observatory, Case Western Reserve University; \facility{\emph{GALEX}}


\begin{thebibliography}{}

\bibitem[Aihara et al.(2011)]{aihara11} Aihara, H., Allende Prieto, C., An, D., et al.\ 2011, \apjs, 193, 29 


\bibitem[Alberts et al.(2011)]{alberts11} Alberts, S., Calzetti, D., Dong, H., et al.\ 2011, \apj, 731, 28 


\bibitem[Astropy Collaboration et al.(2013)]{astropy13} Astropy Collaboration, Robitaille, T.~P., Tollerud, E.~J., et al.\ 2013, \aap, 558, A33 

\bibitem[Bakos et al.(2008)]{bakos08} Bakos, J., Trujillo, I., \& Pohlen, M.\ 2008, \apjl, 683, L103 


\bibitem[Barker et al.(2007)]{barker07} Barker, M.~K., Sarajedini, A., Geisler, D., Harding, P., \& Schommer, R.\ 2007, \aj, 133, 1125 

\bibitem[Barnes et al.(2011)]{barnes11} Barnes, K.~L., van Zee, L., \&
Skillman, E.~D.\ 2011, \apj, 743, 137 

\bibitem[Barnes et al.(2013)]{barnes13} Barnes, K.~L., van Zee, L., \& Dowell, J.~D.\ 2013, \apj, 775, 40 

\bibitem[Beale \& Davies(1969)]{beale69} Beale, J.~S., \& Davies, R.~D.\ 1969, \nat, 221, 531

\bibitem[Bertin \& Arnouts(1996)]{bertin96} Bertin, E., \& Arnouts, S.\ 1996, \aaps, 117, 393 


\bibitem[Bianchi et al.(2003)]{bianchi03} Bianchi, L., Madore, B., Thilker, D., Gil de Paz, A., \& GALEX Science Team 2003, Bulletin of the American Astronomical Society, 35, 91.12 


\bibitem[Bigiel et al.(2008)]{bigiel08} Bigiel, F., Leroy, A., Walter, F., et al.\ 2008, \aj, 136, 2846 


\bibitem[Bigiel et al.(2010)]{bigiel10} Bigiel, F., Leroy, A., Walter, F., et al.\ 2010, \aj, 140, 1194 

\bibitem[Binney \& Merrifield(1998)]{binney98} Binney, J., \& Merrifield, M.\ 1998, Galactic astronomy / James Binney and Michael Merrifield.~ Princeton, NJ : Princeton University Press, 1998.~ (Princeton series in astrophysics)

\bibitem[Boissier et al.(2008)]{boissier08} Boissier, S., Gil de Paz, A., Boselli, A., et al.\ 2008, \apj, 681, 244-257 


\bibitem[Bolatto et al.(2011)]{bolatto11} Bolatto, A.~D., Leroy, A.~K., Jameson, K., et al.\ 2011, \apj, 741, 12 

\bibitem[Bonnell et al.(2004)]{bonnell04} Bonnell, I.~A., Vine, S.~G., \& Bate, M.~R.\ 2004, \mnras, 349, 735

\bibitem[Boselli et al.(2009)]{boselli09} Boselli, A., Boissier, S., Cortese, L., et al.\ 2009, \apj, 706, 1527 


\bibitem[Bottema et al.(1987)]{bottema87} Bottema, R., Shostak, G.~S., \& van der Kruit, P.~C.\ 1987, \nat, 328, 401 

\bibitem[Breiman(1973)]{breiman73} Breiman, L.\ 1973, \emph{Statistics
with a View toward Applications} (Houghton Mifflin, Boston)

\bibitem[Bressan et al.(2012)]{bressan12} Bressan, A., Marigo, P., Girardi, L., et al.\ 2012, \mnras, 427, 127 

\bibitem[Brown et al.(1994)]{brown94} Brown, A.~G.~A., de Geus, E.~J., \& de Zeeuw, P.~T.\ 1994, \aap, 289, 101 

\bibitem[Brown \& Forsythe(1974)]{brown74} Brown, B.~W., \& Forsythe,
A.~B.\ 1974, Robust tests for equality of
variances, \emph{J. Amer. Statist. Assoc.}, 69, 364

\bibitem[Bruzzese et al.(2015)]{bruzzese15} Bruzzese, S.~M., Meurer, G.~R., Lagos, C.~D.~P., et al.\ 2015, \mnras, 447, 618 

\bibitem[Burkholder et al.(2001)]{burkholder01} Burkholder, V., Impey, C., \& Sprayberry, D.\ 2001, \aj, 122, 2318 

\bibitem[Bush et al.(2010)]{bush10} Bush, S.~J., Cox, T.~J., Hayward, C.~C., et al.\ 2010, \apj, 713, 780 

\bibitem[Calzetti et al.(2005)]{calzetti05} Calzetti, D., Kennicutt, R.~C., Jr., Bianchi, L., et al.\ 2005, \apj, 633, 871 


\bibitem[Calzetti(2001)]{calzetti01} Calzetti, D.\ 2001, \pasp, 113, 1449 


\bibitem[Cortese et al.(2006)]{cortese06} Cortese, L., Boselli, A., Buat, V., et al.\ 2006, \apj, 637, 242 


\bibitem[Courtes \& Cruvellier(1961)]{courtes61} Courtes, G., \& Cruvellier, P.\ 1961, Publications of the Observatoire Haute-Provence, 5


\bibitem[Crocker et al.(2015)]{crocker15} Crocker, A.~F., Chandar, R., Calzetti, D., et al.\ 2015, \apj, 808, 76 

\bibitem[Croxall et al.(2016)]{croxall16} Croxall, K.~V., Pogge, R.~W., Berg, D.~A., Skillman, E.~D., \& Moustakas, J.\ 2016, \apj, 830, 4 

\bibitem[da Silva et al.(2014)]{dasilva14} da Silva, R.~L., Fumagalli, M., \& Krumholz, M.~R.\ 2014, \mnras, 444, 3275 

\bibitem[Davidge(2003)]{davidge03} Davidge, T.~J.\ 2003, \aj, 125, 3046 


\bibitem[Davidge(2010)]{davidge10} Davidge, T.~J.\ 2010, \apj, 718, 1428 


\bibitem[de Vaucouleurs et al.(1991)]{devau91} de Vaucouleurs, G., de Vaucouleurs, A., Corwin, H.~G. Jr., Buta,
 R.~J., Paturel, G., \& Fouqu\'e, P. 1991, Third Reference Catalogue
 of Bright Galaxies (Springer(New York))

\bibitem[de Wit et al.(2005)]{dewit05} de Wit, W.~J., Testi, L., Palla, F., \& Zinnecker, H.\ 2005, \aap, 437, 247 

\bibitem[Debattista et al.(2006)]{debattista06} Debattista, V.~P., Mayer, L., Carollo, C.~M., et al.\ 2006, \apj, 645, 209 

\bibitem[Donas et al.(1987)]{donas87} Donas, J., Deharveng, J.~M., Laget, M., Milliard, B., \& Huguenin, D.\ 1987, \aap, 180, 12 

\bibitem[Dopita \& Sutherland(2003)]{dopita03} Dopita, M.~A., \& Sutherland, R.~S.\ 2003, Astrophysics of the diffuse universe, Berlin, New York: Springer, 2003

\bibitem[Dyson \& Williams(1980)]{dyson80} Dyson, J.~E., \& Williams, D.~A.\ 1980, New York, Halsted Press, 1980.~204 p.


\bibitem[Elmegreen \& Hunter(2015)]{elmegreen15} Elmegreen, B.~G., \& Hunter, D.~A.\ 2015, \apj, 805, 145 


\bibitem[Ferguson et al.(1996)]{ferguson96} Ferguson, A.~M.~N., Wyse, R.~F.~G., Gallagher, J.~S., III, \& Hunter, D.~A.\ 1996, \aj, 111, 2265 


\bibitem[Ferguson et al.(1998)]{ferguson98a} Ferguson, A.~M.~N., Gallagher, J.~S., \& Wyse, R.~F.~G.\ 1998, \aj, 116, 673 


\bibitem[Finlator et al.(2000)]{finlator00} Finlator, K., Ivezi{\'c}, {\v Z}., Fan, X., et al.\ 2000, \aj, 120, 2615 


\bibitem[Garc{\'{\i}}a-Ruiz et al.(2002)]{garcia02} Garc{\'{\i}}a-Ruiz, I., Sancisi, R., \& Kuijken, K.\ 2002, \aap, 394, 769 

\bibitem[Garcia-Segura \& Franco(1996)]{garcia96} Garcia-Segura, G., \& Franco, J.\ 1996, \apj, 469, 171 

\bibitem[Gies(1987)]{gies87} Gies, D.~R.\ 1987, \apjs, 64, 545 

\bibitem[Gil de Paz et al.(2005)]{gildepaz05} Gil de Paz, A., Madore, B.~F., Boissier, S., et al.\ 2005, \apjl, 627, L29 


\bibitem[Goddard et al.(2010)]{goddard10} Goddard, Q.~E., Kennicutt, R.~C., \& Ryan-Weber, E.~V.\ 2010, \mnras, 405, 2791 


\bibitem[Greenawalt et al.(1998)]{greenawalt98} Greenawalt, B., Walterbos, R.~A.~M., Thilker, D., \& Hoopes, C.~G.\ 1998, \apj, 506, 135 


\bibitem[Gunawardhana et al.(2011)]{gunawardhana11} Gunawardhana, M.~L.~P., Hopkins, A.~M., Sharp, R.~G., et al.\ 2011, \mnras, 415, 1647 


\bibitem[Haffner et al.(1999)]{haffner99} Haffner, L.~M., Reynolds, R.~J., \& Tufte, S.~L.\ 1999, \apj, 523, 223 

\bibitem[Haffner et al.(2009)]{haffner09} Haffner, L.~M., Dettmar, R.-J., Beckman, J.~E., et al.\ 2009, Reviews of Modern Physics, 81, 969 

\bibitem[Heckman et al.(1995)]{heckman95} Heckman, T., Krolik, J., Meurer, G., et al.\ 1995, \apj, 452, 549 

\bibitem[Helmboldt et al.(2005)]{helmboldt05} Helmboldt, J.~F., Walterbos, R.~A.~M., Bothun, G.~D., \& O'Neil, K.\ 2005, \apj, 630, 824 

\bibitem[Helmboldt et al.(2009)]{helmboldt09} Helmboldt, J.~F., Walterbos, R.~A.~M., Bothun, G.~D., O'Neil, K., \& Oey, M.~S.\ 2009, \mnras, 393, 478 

\bibitem[Hermanowicz et al.(2013)]{hermanowicz13} Hermanowicz, M.~T., Kennicutt, R.~C., \& Eldridge, J.~J.\ 2013, \mnras, 432, 3097 

\bibitem[Hoopes et al.(1996)]{hoopes96} Hoopes, C.~G., Walterbos, R.~A.~M., \& Greenwalt, B.~E.\ 1996, \aj, 112, 1429 

\bibitem[Hoopes et al.(1999)]{hoopes99} Hoopes, C.~G., Walterbos, R.~A.~M., \& Rand, R.~J.\ 1999, \apj, 522, 669 

\bibitem[Hoopes \& Walterbos(2000)]{hoopes00} Hoopes, C.~G., \& Walterbos, R.~A.~M.\ 2000, \apj, 541, 597 

\bibitem[Hoopes et al.(2001)]{hoopes01} Hoopes, C.~G., Walterbos, R.~A.~M., \& Bothun, G.~D.\ 2001, \apj, 559, 878 


\bibitem[Hoversten \& Glazebrook(2008)]{hoversten08} Hoversten, E.~A., \& Glazebrook, K.\ 2008, \apj, 675, 163-187 

\bibitem[Huber(1981)]{huber81} Huber, P.~J.\ 1981, \emph{Robust
 Statistics} (Wiley, New York)

\bibitem[Hunter \& Gallagher(1986)]{hunter86} Hunter, D.~A., \& Gallagher, J.~S., III 1986, \pasp, 98, 5 


\bibitem[Hunter \& Plummer(1996)]{hunter96} Hunter, D.~A., \& Plummer, J.~D.\ 1996, \apj, 462, 732 

\bibitem[Hunter et al.(2003)]{hunter03} Hunter, D.~A., Elmegreen, B.~G., Dupuy, T.~J., \& Mortonson, M.\ 2003, \aj, 126, 1836 

\bibitem[Hunter et al.(2010)]{hunter10} Hunter, D.~A., Elmegreen, B.~G., \& Ludka, B.~C.\ 2010, \aj, 139, 447 

\bibitem[Hunter et al.(2011)]{hunter11} Hunter, D.~A., Elmegreen, B.~G., Oh, S.-H., et al.\ 2011, \aj, 142, 121 

\bibitem[Hunter(2007)]{hunter07}
 Hunter, J.~D. 2007, \emph{Computing in Science \& Engineering}, 9, 90


\bibitem[Kellar et al.(2012)]{kellar12} Kellar, J.~A., Salzer, J.~J., Wegner, G., Gronwall, C., \& Williams, A.\ 2012, \aj, 143, 145 

\bibitem[Kenney et al.(1991)]{kenney91} Kenney, J.~D.~P., Scoville, N.~Z., \& Wilson, C.~D.\ 1991, \apj, 366, 432 

\bibitem[Kennicutt(1983)]{kennicutt83} Kennicutt, R.~C., Jr.\ 1983, \apj, 272, 54 

\bibitem[Kennicutt(1989)]{kennicutt89} Kennicutt, R.~C., Jr.\ 1989, \apj, 344, 685 

\bibitem[Kennicutt et al.(1994)]{kennicutt94} Kennicutt, R.~C., Jr., Tamblyn, P., \& Congdon, C.~E.\ 1994, \apj, 435, 22 

\bibitem[Kennicutt(1998)]{kennicutt98} Kennicutt, R.~C., Jr.\ 1998, \apj, 498, 541 

\bibitem[Kennicutt et al.(2007)]{kennicutt07} Kennicutt, R.~C., Jr., Calzetti, D., Walter, F., et al.\ 2007, \apj, 671, 333 


\bibitem[Kennicutt et al.(2008)]{kennicutt08} Kennicutt, R.~C., Jr., Lee, J.~C., Funes, J.~G., et al.\ 2008, \apjs, 178, 247-279 

\bibitem[Kennicutt \& Evans(2012)]{kennicutt12} Kennicutt, R.~C., \& Evans, N.~J.\ 2012, \araa, 50, 531 


\bibitem[Kornreich et al.(1998)]{kornreich98} Kornreich, D.~A., Haynes, M.~P., \& Lovelace, R.~V.~E.\ 1998, \aj, 116, 2154 

\bibitem[Koyama et al.(2015)]{koyama15} Koyama, Y., Kodama, T., Hayashi, M., et al.\ 2015, \mnras, 453, 879 


\bibitem[Kroupa(2001)]{kroupa01} Kroupa, P.\ 2001, \mnras, 322, 231 

\bibitem[Lada \& Lada(2003)]{lada03} Lada, C.~J., \& Lada, E.~A.\ 2003, \araa, 41, 57 

\bibitem[Laine et al.(2016)]{laine16} Laine, J., Laurikainen, E., \& Salo, H.\ 2016, \aap, 596, A25 

\bibitem[Lamb et al.(2010)]{lamb10} Lamb, J.~B., Oey, M.~S., Werk, J.~K., \& Ingleby, L.~D.\ 2010, \apj, 725, 1886 

\bibitem[Larson(1973)]{larson73} Larson, R.~B.\ 1973, \mnras, 161, 133 

\bibitem[Lee et al.(2009)]{lee09} Lee, J.~C., Gil de Paz, A., Tremonti, C., et al.\ 2009, \apj, 706, 599-613 


\bibitem[Lee et al.(2011)]{lee11} Lee, J.~C., Gil de Paz, A., Kennicutt, R.~C., Jr., et al.\ 2011, \apjs, 192, 6 


\bibitem[Leitherer et al.(1999)]{leitherer99} Leitherer, C., Schaerer, D., Goldader, J.~D., et al.\ 1999, \apjs, 123, 3 

\bibitem[Lemonias et al.(2011)]{lemonias11} Lemonias, J.~J., Schiminovich, D., Thilker, D., et al.\ 2011, \apj, 733, 74  

\bibitem[Leroy et al.(2012)]{leroy12} Leroy, A.~K., Bigiel, F., de Blok, W.~J.~G., et al.\ 2012, \aj, 144, 3 

\bibitem[Levene(1960)]{levene60} Levene, H.\ 1960, Robust tests for
 equality of variances.  In \emph{Contributions to Probability and
 Statistics} (I. Olkin, ed.) 278-292, Stanford Univ. Press, Palo
 Alto, CA

\bibitem[Lim \& Loh(1996)]{limloh96} Lim, T.~S., \& Loh, W.~Y.\
 1996, \emph{Computational Statistics \& Data Analysis}, 22, 287

\bibitem[Madsen et al.(2006)]{madsen06} Madsen, G.~J., Reynolds, R.~J., \& Haffner, L.~M.\ 2006, \apj, 652, 401 


\bibitem[Martin \& Kennicutt(2001)]{martin01} Martin, C.~L., \& Kennicutt, R.~C., Jr.\ 2001, \apj, 555, 301 


\bibitem[Massey \& Foltz(2000)]{massey00} Massey, P., \& Foltz, C.~B.\ 2000, \pasp, 112, 566 


\bibitem[Massey et al.(1988)]{massey88} Massey, P., Strobel, K., Barnes, J.~V., \& Anderson, E.\ 1988, \apj, 328, 315 


\bibitem[McGaugh \& Bothun(1994)]{mcgaugh94a} McGaugh, S.~S., \& Bothun, G.~D.\ 1994, \aj, 107, 530 


\bibitem[McGaugh \& de Blok(1997)]{mcgaugh97} McGaugh, S.~S., \& de Blok, W.~J.~G.\ 1997, \apj, 481, 689 


\bibitem[Meidt et al.(2009)]{meidt09} Meidt, S.~E., Rand, R.~J., \& Merrifield, M.~R.\ 2009, \apj, 702, 277 

\bibitem[Meurer et al.(1995)]{meurer95} Meurer, G.~R., Heckman, T.~M., Leitherer, C., et al.\ 1995, \aj, 110, 2665 

\bibitem[Meurer et al.(1999)]{meurer99} Meurer, G.~R., Heckman, T.~M., \& Calzetti, D.\ 1999, \apj, 521, 64 


\bibitem[Meurer et al.(2009)]{meurer09} Meurer, G.~R., Wong, O.~I., Kim, J.~H., et al.\ 2009, \apj, 695, 765 

\bibitem[Meurer(2017)]{meurer17} Meurer, G.~R.\ 2017, Formation and Evolution of Galaxy Outskirts, 321, 172 


\bibitem[Mihos et al.(2012)]{mihos12} Mihos, J.~C., Keating, K.~M., Holley-Bockelmann, K., Pisano, D.~J., \& Kassim, N.~E.\ 2012, \apj, 761, 186 

\bibitem[Mihos et al.(2013)]{mihos13} Mihos, J.~C., Harding, P., Spengler, C.~E., Rudick, C.~S., \& Feldmeier, J.~J.\ 2013, \apj, 762, 82 


\bibitem[Mihos et al.(2017)]{mihos17} Mihos, J.~C., Harding, P., Feldmeier, J.~J., et al.\ 2017, \apj, 834, 16 


\bibitem[Minchev et al.(2011)]{minchev11} Minchev, I., Famaey, B., Combes, F., et al.\ 2011, \aap, 527, A147 

\bibitem[Moffat et al.(1998)]{moffat98} Moffat, A.~F.~J., Marchenko,
S.~V., Seggewiss, W., et al.\ 1998, \aap, 331, 949

\bibitem[Morrison et al.(1990)]{morrison90} Morrison, H.~L., Flynn, C., \& Freeman, K.~C.\ 1990, \aj, 100, 1191 


\bibitem[Morrissey et al.(2007)]{morrissey07} Morrissey, P., Conrow, T., Barlow, T.~A., et al.\ 2007, \apjs, 173, 682 

\bibitem[Oey et al.(2013)]{oey13} Oey, M.~S., Lamb, J.~B., Kushner, C.~T., Pellegrini, E.~W., \& Graus, A.~S.\ 2013, \apj, 768, 66 

\bibitem[Oliphant(2007)]{oliphant07}
 Oliphant, T.~E. 2007, \emph{Computing in Science \& Engineering}, 9,
 10

\bibitem[Pflamm-Altenburg \& Kroupa(2008)]{pflamm08} Pflamm-Altenburg, J., \& Kroupa, P.\ 2008, \nat, 455, 641 

\bibitem[Pflamm-Altenburg et al.(2009)]{pflamm09} Pflamm-Altenburg,
 J., Weidner, C., \& Kroupa, P.\ 2009, \mnras, 395, 394

\bibitem[Pflamm-Altenburg et al.(2013)]{pflamm13} Pflamm-Altenburg, J., Gonz{\'a}lez-L{\'o}pezlira, R.~A., \& Kroupa, P.\ 2013, \mnras, 435, 2604 

\bibitem[Pilyugin et al.(2014)]{pilyugin14} Pilyugin, L.~S., Grebel, E.~K., \& Kniazev, A.~Y.\ 2014, \aj, 147, 131 


\bibitem[Quireza et al.(2006)]{quireza06} Quireza, C., Rood, R.~T., Bania, T.~M., Balser, D.~S., \& Maciel, W.~J.\ 2006, \apj, 653, 1226 


\bibitem[Rand(1996)]{rand96} Rand, R.~J.\ 1996, \apj, 462, 712 


\bibitem[Rand(1997)]{rand97} Rand, R.~J.\ 1997, \apj, 474, 129 


\bibitem[Rela{\~n}o et al.(2012)]{relano12} Rela{\~n}o, M., Kennicutt, R.~C., Jr., Eldridge, J.~J., Lee, J.~C., \& Verley, S.\ 2012, \mnras, 423, 2933 


\bibitem[Reynolds et al.(1977)]{reynolds77} Reynolds, R.~J., Roesler, F.~L., \& Scherb, F.\ 1977, \apj, 211, 115 


\bibitem[Reynolds(1985)]{reynolds85} Reynolds, R.~J.\ 1985, \apjl, 298, L27 


\bibitem[Reynolds(1990)]{reynolds90} Reynolds, R.~J.\ 1990, The Galactic and Extragalactic Background Radiation, 139, 157 


\bibitem[Ro{\v s}kar et al.(2008)]{roskar08} Ro{\v s}kar, R., Debattista, V.~P., Quinn, T.~R., Stinson, G.~S., \& Wadsley, J.\ 2008, \apjl, 684, L79 


\bibitem[Ro{\v s}kar et al.(2012)]{roskar12} Ro{\v s}kar, R., Debattista, V.~P., Quinn, T.~R., \& Wadsley, J.\ 2012, \mnras, 426, 2089 

\bibitem[Rownd et al.(1994)]{rownd94} Rownd, B.~K., Dickey, J.~M., \& Helou, G.\ 1994, \aj, 108, 1638

\bibitem[Rudick et al.(2010)]{rudick10} Rudick, C.~S., Mihos, J.~C., Harding, P., et al.\ 2010, \apj, 720, 569 

\bibitem[S{\'a}nchez-Bl{\'a}zquez et al.(2009)]{sanchez09} S{\'a}nchez-Bl{\'a}zquez, P., Courty, S., Gibson, B.~K., \& Brook, C.~B.\ 2009, \mnras, 398, 591 

\bibitem[Sancisi(1976)]{sancisi76} Sancisi, R.\ 1976, \aap, 53, 159 

\bibitem[Schlafly \& Finkbeiner(2011)]{schlafly11}
 Schlafly, E.~F., \& Finkbeiner, D.~P. 2011, \apj, 737, 103

\bibitem[Schlegel et al.(1998)]{schlegel98}
 Schlegel, D.~J., Finkbeiner, D.~P., \& Davis, M. 1998, \apj, 500, 525


\bibitem[Schmidt(1959)]{schmidt59} Schmidt, M.\ 1959, \apj, 129, 243 


\bibitem[Schombert et al.(2013)]{schombert13} Schombert, J., McGaugh, S., \& Maciel, T.\ 2013, \aj, 146, 41 

\bibitem[Sch{\"o}nrich \& Binney(2009)]{schonrich09} Sch{\"o}nrich, R., \& Binney, J.\ 2009, \mnras, 399, 1145 

\bibitem[Schruba et al.(2011)]{schruba11} Schruba, A., Leroy, A.~K., Walter, F., et al.\ 2011, \aj, 142, 37 


\bibitem[Scowen et al.(1992)]{scowen92} Scowen, P.~A., Dufour, R.~J., \& Hester, J.~J.\ 1992, \aj, 104, 92 


\bibitem[Seibert et al.(2005)]{seibert05} Seibert, M., Martin, D.~C., Heckman, T.~M., et al.\ 2005, \apjl, 619, L55 


\bibitem[Sellwood \& Binney(2002)]{sellwood02} Sellwood, J.~A., \& Binney, J.~J.\ 2002, \mnras, 336, 785 


\bibitem[Silk \& Werner(1969)]{silk69} Silk, J., \& Werner, M.~W.\ 1969, \apj, 158, 185 


\bibitem[Skillman(1987)]{skillman87} Skillman, E.~D.\ 1987, NASA Conference Publication, 2466

\bibitem[Slater et al.(2009)]{slater09} Slater, C.~T., Harding, P., \& Mihos, J.~C.\ 2009, \pasp, 121, 1267 

\bibitem[Spitzer \& Jenkins(1975)]{spitzer75} Spitzer, L., Jr., \& Jenkins, E.~B.\ 1975, \araa, 13, 133 


\bibitem[Spitzer \& Tomasko(1968)]{spitzer68} Spitzer, L., Jr., \& Tomasko, M.~G.\ 1968, \apj, 152, 971 

\bibitem[Sullivan et al.(2004)]{sullivan04} Sullivan, M., Treyer, M.~A., Ellis, R.~S., \& Mobasher, B.\ 2004, \mnras, 350, 21 

\bibitem[Thilker et al.(2002)]{thilker02} Thilker, D.~A., Walterbos, R.~A.~M., Braun, R., \& Hoopes, C.~G.\ 2002, \aj, 124, 3118 


\bibitem[Thilker et al.(2005)]{thilker05} Thilker, D.~A., Bianchi, L., Boissier, S., et al.\ 2005, \apjl, 619, L79 


\bibitem[Thilker et al.(2007)]{thilker07} Thilker, D.~A., Bianchi, L., Meurer, G., et al.\ 2007, \apjs, 173, 538 

\bibitem[Toomre(1964)]{toomre64} Toomre, A.\ 1964, \apj, 139, 1217 

\bibitem[Torres-Peimbert et al.(1974)]{torres74} Torres-Peimbert, S., Lazcano-Araujo, A., \& Peimbert, M.\ 1974, \apj, 191, 401 


\bibitem[van der Hulst \& Huchtmeier(1979)]{vanderhulst79} van der Hulst, J.~M., \& Huchtmeier, W.~K.\ 1979, \aap, 78, 82 

\bibitem[van der Hulst et al.(1987)]{vanderhulst87} van der Hulst, J.~M., Skillman, E.~D., Kennicutt, R.~C., \& Bothun, G.~D.\ 1987, \aap, 177, 63 

\bibitem[van der Hulst et al.(2001)]{vanderhulst01}
 van der Hulst, J.~M., van Albada, T.~S., \& Sancisi,
 R. 2001, \emph{ASP Conf. Ser. 240, Gas and Galaxy Evolution},
 ed. J.E. Hibbard, M. Rupen, \& J.H. van Gorkum (San Francisco, CA:
 ASP), 451

\bibitem[van der Kruit(1987)]{vanderkruit87} van der Kruit, P.~C.\ 1987, \aap, 173, 59 

\bibitem[VanderPlas et al.(2012)]{vanderplas12} VanderPlas, J., Connolly, A.~J., Ivezic, Z., \& Gray, A.\ 2012, Proceedings of Conference on Intelligent Data Understanding (CIDU), pp.~47-54, 2012., 47 

\bibitem[Vandervoort(1970)]{vandervoort70} Vandervoort, P.~O.\ 1970, \apj, 161, 87 

\bibitem[van Eymeren et al.(2011)]{vaneymeren11} van Eymeren, J., J{\"u}tte, E., Jog, C.~J., Stein, Y., \& Dettmar, R.-J.\ 2011, \aap, 530, A30 


\bibitem[van Zee et al.(1997)]{vanzee97} van Zee, L., Haynes, M.~P., Salzer, J.~J., \& Broeils, A.~H.\ 1997, \aj, 113, 1618 


\bibitem[Vlaji{\'c} et al.(2009)]{vlajic09} Vlaji{\'c}, M., Bland-Hawthorn, J., \& Freeman, K.~C.\ 2009, \apj, 697, 361 


\bibitem[Vlaji{\'c} et al.(2011)]{vlajic11} Vlaji{\'c}, M., Bland-Hawthorn, J., \& Freeman, K.~C.\ 2011, \apj, 732, 7 

\bibitem[Vogel et al.(1995)]{vogel95} Vogel, S.~N., Weymann, R., Rauch, M., \& Hamilton, T.\ 1995, \apj, 441, 162 

\bibitem[Waller et al.(1997)]{waller97} Waller, W.~H., Bohlin, R.~C., Cornett, R.~H., et al.\ 1997, \apj, 481, 169 

\bibitem[Walter et al.(2008)]{walter08} Walter, F., Brinks, E., de Blok, W.~J.~G., et al.\ 2008, \aj, 136, 2563-2647 

\bibitem[Walterbos \& Braun(1994)]{walterbos94} Walterbos, R.~A.~M., \& Braun, R.\ 1994, \apj, 431, 156 


\bibitem[Watkins et al.(2014)]{watkins14} Watkins, A.~E., Mihos,
 J.~C., Harding, P., \& Feldmeier, J.~J.\ 2014, \apj, 791, 38 


\bibitem[Watkins et al.(2016)]{watkins16} Watkins, A.~E., Mihos, J.~C., \& Harding, P.\ 2016, \apj, 826, 59 

\bibitem[Weidner et al.(2004)]{weidner04} Weidner, C., Kroupa, P., \& Larsen, S.~S.\ 2004, \mnras, 350, 1503 

\bibitem[Weisz et al.(2012)]{weisz12} Weisz, D.~R., Johnson, B.~D., Johnson, L.~C., et al.\ 2012, \apj, 744, 44 

\bibitem[Whitmore et al.(2011)]{whitmore11} Whitmore, B.~C., Chandar, R., Kim, H., et al.\ 2011, \apj, 729, 78 

\bibitem[Wood \& Reynolds(1999)]{wood99} Wood, K., \& Reynolds, R.~J.\ 1999, \apj, 525, 799 


\bibitem[Zaritsky \& Christlein(2007)]{zaritsky07} Zaritsky, D., \& Christlein, D.\ 2007, \aj, 134, 135 


\bibitem[Zasov \& Simakov(1988)]{zasov88} Zasov, A.~V., \& Simakov, S.~G.\ 1988, Astrophysics, 29, 518 


\bibitem[Zheng et al.(2015)]{zheng15} Zheng, Z., Thilker, D.~A., Heckman, T.~M., et al.\ 2015, \apj, 800, 120 


\bibitem[Zurita et al.(2002)]{zurita02} Zurita, A., Beckman, J.~E., Rozas, M., \& Ryder, S.\ 2002, \aap, 386, 801 


\end{thebibliography}
\end{document}